\documentclass[12pt,a4paper,oneside]{report}

\usepackage[T1]{fontenc}
\usepackage[utf8]{inputenc}
\usepackage[english]{babel}
\usepackage[left=1.5cm, right=3cm, top=1.5cm, bottom=2cm, bindingoffset=1.5cm, head=15pt]{geometry} 
\usepackage{setspace}
\usepackage{tocloft}
\usepackage{amsmath}
\usepackage{amsthm}
\usepackage{amssymb}
\usepackage{amsfonts}
\usepackage{graphicx}
\usepackage[all]{xy}
\usepackage{multicol}
\usepackage{longtable}
\usepackage{float}
\usepackage{anysize}
\usepackage{appendix}
\usepackage{lscape} 
\usepackage{pdflscape}
\usepackage{multirow}
\usepackage{listings}
\usepackage{color}
\usepackage{setspace}
\usepackage{enumerate} 
\usepackage{stackrel}
\usepackage{MnSymbol}
\usepackage{mathtools}

\newtheorem{theorem}{Theorem}[section]
\newtheorem{corollary}{Corollary}

\newtheorem{proposition}{Proposition}
 
\theoremstyle{definition}
\newtheorem{definition}{Definition}[section]

\usepackage[backref=page]{hyperref}
\hypersetup{
    colorlinks=true,
    allcolors=blue,
}

\begin{document}



\marginsize{3.0cm}{3.0cm}{4.0cm}{3.0cm}
\renewcommand*{\contentsname}{INDEX}
\renewcommand*{\listtablename}{Tables index}
\renewcommand*{\listfigurename}{Figures Index}
\renewcommand{\baselinestretch}{1.0}
\renewcommand{\appendixname}{Anexos}
\renewcommand{\appendixtocname}{Anexos}
\renewcommand{\appendixpagename}{Anexos}
\renewcommand{\thetable}{\arabic{chapter}.\arabic{table}}
\renewcommand*{\tablename}{Table}
\renewcommand*{\chaptername}{Chapter}
\renewcommand*{\thechapter}{\Roman{chapter}}
\renewcommand{\thesection}{\arabic{chapter}.\arabic{section}}
\renewcommand{\figurename}{Figure}
\renewcommand{\thefigure}{\arabic{chapter}.\arabic{figure}}
\renewcommand{\theequation}{\arabic{chapter}.\arabic{equation}}


\begin{titlepage}

\begin{center}
    UNIVERSITY OF GUADALAJARA \\

    \vspace{0.5cm}
    
    \begin{Large}
    {\bf
    Ontology of the Theory of Relativity
    }
    \end{Large}
\end{center}
\begin{center}
{\large \bf Salvador D. Escobedo}\footnote{salvador.escobedo@alumno.udg.mx} \\

\vspace{0.5cm}

Advisor: \textbf{Dr. José Edgar Madriz Aguilar}\footnote{jose.madriz@academicos.udg.mx}

Co-advisor: \textbf{Dr. Américo Peraza Álvarez}\footnote{americo.peraza@academicos.udg.mx}

\vspace{0.5cm}

{\it Bachelor Thesis in Physics.}\footnote{This is a translation from the original thesis in Spanish \url{https://arxiv.org/abs/2302.14809v1}.}
\end{center}

\begin{small}
\textbf{Abstract}: Both relativistic mechanics and Newtonian mechanics are based on principles that have ontological implications. We propose a series of formalisms that rigorously define the ontology underlying mechanical theories, in order to clarify and formally establish the ontology of the physics of motion. Special attention has been paid to relativistic theories. Through the proposed methodology, the concept of ontological consistency is developed and the conditions required for such consistency to be satisfied in any theory are established. In particular, the consistency test is performed for Newtonian mechanics, special relativity theory, and general relativity theory.  
\end{small}

\vspace{1cm}

\vspace{0.1cm}

\begin{flushright}
Spring 2020
\end{flushright}

\end{titlepage}

\newpage

\begin{titlepage}

\begin{flushright}
{\large \bf DEDICATION}
\\
\textit{To my parents, for their unconditional support.}
\\
\textit{To my fiancée, for her constant companionship.}

\end{flushright}
\end{titlepage}


\begin{titlepage}

\begin{flushright}
{\large \bf ACKNOWLEDGMENTS}
\end{flushright}

I want to express my gratitude to my parents for their unwavering support, without which pursuing a degree in physics would not have been possible.\\

To my fiancée, Adriana Duarte, for being a steadfast support throughout my academic journey.\\

To my thesis advisor, Dr. José Edgar Madriz Aguilar, for believing in this project and for his personal and academic guidance. \\

To my co-advisor, Dr. Américo Peraza Álvarez, for the trust he placed in me.\\

To Dr. Mariana Sarahí Montes Navarro, for her continuous assistance and academic advice. \bigskip

\begin{flushright}
Salvador Daniel Escobedo Casillas.
\end{flushright}

\end{titlepage}


\clearpage
\tableofcontents
\cleardoublepage
\newpage


\chapter*{Introduction} 
\addcontentsline{toc}{chapter}{Introduction} 
\markboth{Introduction}{Introduction} 

\qquad Ontology is a branch of philosophy, which has been, and for a long time, a natural starting point in scientific inquiry. Despite the fact that many scientists do not realize it, many of the fundamental propositions of modern science are not actually scientific propositions, but assertions of ontological nature. Let us remember that ontology –also referred to as metaphysics\footnote{Some authors distinguish between metaphysics and ontology, treating ontology as a type of general metaphysics, and therefore as a part of metaphysics. <<Other authors, on the other hand, consider this division deplorable, as it breaks the unity of the investigation of being (\textit{esse}), the theme of metaphysics and ontology or, if you will, of \textit{metaphysics-ontology}>> (Ferrater Mora, J. (1994) Dictionary of Philosophy, entry Ontology). We will not follow this distinction either, as it is irrelevant to our purposes.}– is the branch of philosophy that deals with the study of being, and therefore when we talk about ontology in physics, we refer to the study of material realities that are independent of our mathematical descriptions and our \textit{frames of reference} \cite{3}. Thus, for example, a physicist can describe electromagnetic interactions using a certain mathematical object called a \textit{field}; the Maxwell's equations, as beautiful and elegantly as they govern interactions in this field, do not provide us with information about the real existence of the field itself. In other words, they do not inherently establish whether the electric field is a real entity or a mere mathematical tool. This type of problem has given rise to an epistemological debate whose two main opposing currents are realism and instrumentalism.\\

It is also noteworthy that significant results in contemporary physics have been pro\-posed and discussed in the past by some Greek philosophers, who were strangers to any kind of advanced experimentation \cite{5,2}. The assertion, for instance, that matter is composed of atoms was not first proposed by Dalton in the 19th century, but by Democritus in 400 B.C.\\

Nevertheless, the usefulness of ontology within science has been diffusely discussed. Certainly, it is a task that philosophers can engage with, but not everyone agrees when it comes to determining whether the study of it would benefit physicists. This was at least the case until the year 1900, which marks the beginning of modern physics with the emergence of new theories that have challenged our \textit{intuition} regarding the physical world \cite{4,6}. Suddenly, the term 'ontology' has become common in efforts to elucidate quantum mechanics.

Putting it plainly and simply, ontology would be useful in physics when we can derive experimentally verifiable propositions from it and when it helps us understand the \textit{whys} of theories that only establish quantitative relationships through equations. Thus, Newton's physics establishes the law of universal gravitation, defining that all bodies attract each other with a force whose magnitude depends on the masses of the bodies attracting each other and the distance between them. Despite the great advancement brought about by this discovery, it leaves a question in the shadows: Why do these bodies attract each other? 

The general theory of relativity proposed a more satisfactory explanation: it explained that gravity is actually the effect of a space-time curvature created around massive bodies. Thus, Einstein provided a formal answer to the problem of what gravity is and why it occurs, but he relied on assumptions that are far from being evident, such as the exotic fact that the speed of light is constant for any observer \cite{1}. 

But wouldn't it be of some utility for physics to deeply understand why the speed of light is invariant, instead of just accepting it as a hypothesis? Certainly. Because from a theory from which we can deduce the principles of relativity, more principles could be deduced that might give rise to new physical theories, all of which have the potential to be experimentally verified.

Similarly, it would be of great utility for physical science to understand the \textit{why}\footnote{It is often said that science doesn't concern itself with the \textit{why}, but with the \textit{how}. However, it's important to note that \textit{why} and \textit{how} are not technical terms, but rather words from natural language, and therefore sensitive to context. Because of this, \textit{why} and \textit{how} are often equivalent in many cases.} of the fundamental principles of quantum mechanics. A satisfactory interpretation that accounts for the success of the Schrödinger equation would not be negligible. The deeper our understanding of nature, the more universal the general postulates of our theories will be and the greater their predictive power.
Hence arises the effort to unify the two great branches of modern physics, relativity and quantum mechanics, into a deeper theory that encompasses both and explains to us, with the greatest generality possible, how the universe functions.

In relation to this, in the late 60s an interesting proposal called the \textit{String theory} emerged, which manages to unify all of physics. The string theory starts from the assumption that everything is made of objects called \textit{strings}, whose vibrations give rise to all particles in the universe (note that this is an ontological proposition). From this ontology and certain mathematical relationships, all known physical laws are deduced. Not one more\footnote{That can be subject to experimental verification.}. This is one of the problems with this theory: its predictive power does not surpass that of existing theories, and it doesn't offer anything new that can be experimentally tested \cite{hedrich2007internal}. The use of an equation as a starting point brings us back to the issue that had already arisen with Newton and universal gravitation; it tells us how much and with what magnitude, but it doesn't explain the causes.

Currently, physics has reached a point where the more general and powerful theories become, the less intelligible they are. We can find solutions to the Schrödinger equation, but we don't know where it comes from or based on what principles it operates\footnote{We don't know \textit{based on what principles it operates}, meaning that the underlying mechanisms that would allow for a unambiguous interpretation of the mathematical formalism are still a matter of debate.}. Wouldn't it be more convenient if we could create an ontological theory that explains the causal relationships and from which we can extract the mathematical equations, rather than just starting from an equation and giving up on investigating, claiming that the formula works without truly understanding the \textit{why}?

This question, which obviously we have not been the first to raise, has been a recurring inquiry in science, especially since the year 1900, and has interested both philosophers of science and physicists. Werner Heisenberg himself, dissatisfied with the purely mathematical explanation of physical reality, said: \begin{quote}This is precisely my difficulty ... that I don't know what more to demand. But I feel, in a certain way, cheated by the logic with which this mathematical framework functions...\cite{7}\end{quote}

A more vehement and comprehensive defense of the need to address the principles of physics through ontological concepts can be found in the work of Mario Bunge\footnote{See, for example, \cite{bunge2011ontologia} or \cite{bunge1973filosofia}, also consult \cite{Azz}}. In this thesis, we propose to develop an ontology that can account for the laws of physics. However, encompassing all of them proves to be a project too ambitious and, for now, beyond our reach. 
For this reason, we will limit ourselves to ontologically grounding only classical mechanics and special and general relativity –which, in our view, is no small task–. The reason for starting by grounding classical mechanics is that many of its postulates find a natural approach to the theory of relativity through this means, while also serving as the first step in the ontological study of a theory that is already profound and complex.

To achieve these stated objectives, the thesis has been structured as follows:

Firstly, we will develop a rigorous formalization of physics. The aim of this chapter is to rigorously define the concept of \textit{ontology}, as well as its relation to physics and its empirical part. We will precisely define what we mean when we talk about a \textit{physical theory}, a \textit{formal model}, an \textit{empirical interpretation}, and an \textit{ontological interpretation}.

Next, we will define the most fundamental notions of our formalism. In this chapter, we will start with the notion of change and from there address the problem of making sense of concepts such as space, trajectory, displacement, distance, duration, simultaneity, and time. These are all concepts of extreme importance within any physical theory.

Subsequently, using the principles established in the previous chapter, we will develop a methodology that allows us to test current physical theories in such a way that we can verify if they fulfill the condition of ontological consistency. This is where the approach we are proposing reveals its usefulness, as this is when we arrive at an understanding of the principles of relativity completely grounded in an ontological theory.

\chapter{Formalization of Physics}
\section{Fundamental Definitions}
\qquad As previously stated, ontology studies being in the most general manner possible. However, keep in mind that when we say \textit{being}, we refer to everything endowed with real existence; to be a being implies existence, and existence in turn implies to be a being. A being that does not exist would not be a being, but rather a non-being. For this reason, ontology always refers to reality, not abstract objects whose existence could only be established in our minds.
So, when we say that a property of something is \textit{ontological}, we mean that it is truly real and not just a logical or verbal attribute.

\begin{definition}
We define the set of all real beings as the set
\begin{equation}
     \textbf{E} = \{x : x \textnormal{ is ontologically real}\} 
\end{equation}
\end{definition}\medskip

However, real objects are not accessible to us except through the senses. This is particularly important in physics, as we do not access material reality without them \cite{Loka}.
Thus, it will also be convenient to define the set of all \textit{empirical phenomena}, namely those objects that are perceptible through the senses.

\begin{definition}
We define the set of empirical objects as the set

\begin{equation}
     \textbf{S} = \{p : p \textnormal{ is a property accessible to the senses}\} 
\end{equation}
\label{def:sens}
\end{definition}

The fact that we have defined the elements of \textbf{S} as properties is not arbitrary. This is because our sensory perception is incapable of directly perceiving physical objects\footnote{Our senses do not perceive substances, only accidents.}, but only through the effects they produce on our sensory organs or measurement instruments. These effects, as perceived by our consciousness, are what we call \textit{phenomena}.


A \textbf{model} $\mathcal{M}$ is a structure that contains terms, propositions, and also a collection of signs that relate the terms and propositions among themselves (functors, relators, predicates, etc.).

Formally, a model $\mathcal{M}$ can be defined as a triple:
\begin{equation}
    \mathcal{M} = (T,P,D)
\end{equation}

Where $T$ is a collection of terms, $P$ a collection of propositions, and $D$ a collection of relational signs (functors, relators, predicates, etc.)

\paragraph{Notation:}
If $t$ is an element of $T$ of a model $\mathcal{M}$, for simplicity we will write $t \in \mathcal{M}$ instead of <<$t \in T$ where $T$ is the collection of terms of $\mathcal{M}$>>. The same will be done with the elements of $P$ and $D$.
\\

If the collection of propositions $P$ can be divided into two disjoint sets $A$ and $B$, with $P=A\cup B$, such that the propositions in $B$ are logical consequences of the propositions in $A$, which are called \textit{axioms}, then we say that the model $\mathcal{M}$ is an \textbf{axiomatic theory}.

Now, let $U$ be any set. We say that $\mathcal{M}$ is a \textbf{model of} $U$ when we add some rule that associates the terms and propositions of $\mathcal{M}$ with the objects and facts of $U$, respectively. The set $U$ is called the \textbf{universe of discourse} of $\mathcal{M}$, and the said rules of association are referred to as \textbf{interpretations}.

\begin{definition}
A proposition $p$ of a model $\mathcal{M}$ is \textbf{true} under an interpretation $I$ if and only if the fact signified by proposition $p$, under interpretation $I$, does indeed occur in the universe $U$.
\end{definition}\medskip

\begin{definition}
We say that a model $\mathcal{M}$ is \textbf{valid} with respect to an interpretation $J$, when every proposition $p$ of $\mathcal{M}$ interpreted by $J$ is true.
\end{definition}


\section{The Physical Model and its Interpretations}

\qquad Let $\mathcal{M}$ be a formal model, and let $\textbf{S}$ be the set of measurable and observable empirical phenomena (def. \ref{def:sens}). An \textbf{empirical interpretation} $I$ is a relation that associates the properties of objects in the model $\mathcal{M}$ with the phenomena in $\textbf{S}$, assigning empirical and observable reality to certain values in $\mathcal{M}$.
On the other hand, an \textbf{ontological interpretation} $O$ is an association between the facts and objects of the formal model $\mathcal{M}$ and the facts and objects of the real universe $\textbf{E}$
\begin{displaymath}
    \xymatrix{
		       & \mathcal{M} \ar[dr]^I \ar[r]^O & \textbf{E} \ar[d]^F \ar[l]\\
         & & \textbf{S}\ar[u]\ar[ul]}
\end{displaymath}

We call the relation $F$ the \textbf{physical interpretation}, and it is the composition of the empirical interpretation $I$ and the ontological interpretation $O$, that is
\begin{equation}
    F = I \circ O
\end{equation}

We say that a \textbf{physical theory} $\mathcal{F}$ is a triad composed of a formal structure or model $\mathcal{M}$, an empirical interpretation $I$, and an ontological interpretation $O$ that assigns reality to the objects in $\mathcal{M}$. In symbols:
\begin{equation}
    \mathcal{F} = \left(\mathcal{M}, I, O \right)
\end{equation}
At times, we will refer to the model $\mathcal{M}$ as a \textit{physical model} when it is part of a physical theory $\mathcal{F}$.

As can be observed, this structure reflects the natural way in which human knowledge proceeds. $\textbf{S}$ represents all those facts and properties that are directly accessible to our senses or measurement instruments. The model $\mathcal{M}$ represents the product of reasoning obtained through abstraction, which combined with $\textbf{S}$ allows us to have an approximation of the world as it is, represented by $\textbf{E}$.

For this to be possible, there must exist a certain set of objective rules or laws, such that not every proposition is true in the described universe. In other words, not every model or interpretation will be true in $\textbf{E}$ or $\textbf{S}$.

To this set of laws regarding $\textbf{E}$, we refer to as \textbf{Ontology}; the fact that an interpretation of a model satisfies an ontology is referred to as \textit{ontological consistency}.

The set of laws that determine the truth of a model with respect to human thought is called \textbf{logic}. The fact that a model satisfies such laws is referred to as \textit{logical consistency}.

Likewise, the set of sensible properties of $\textbf{S}$ that we have at our disposal is called \textbf{empirical data}, and when the interpretation of a given model is consistent with the empirical data, we say that the model \textit{fits the experience}.

In summary, the conditions for the validity of a physical model are:
\begin{itemize}
    \item Regarding thought: The model $\mathcal{M}$ must possess \textit{logical consistency}.
    \item Regarding $O$: The model $\mathcal{M}$ must satisfy the ontological principles. \textit{Ontological consistency}
    \item Regarding $I$: The model $\mathcal{M}$ must satisfy the empirical evidence. \textit{Empirical consistency}
\end{itemize}
Formally, this is defined as follows:

\begin{definition}
A physical theory $\mathcal{F}$, with a logically consistent model $\mathcal{M}$, is valid if and only if the propositions of $\mathcal{M}$ interpreted by $O$ are true in $\textbf{E}$, and the propositions of $\mathcal{M}$ interpreted by $I$ are true in $\textbf{S}$.
\end{definition}

The following principle establishes an immediate extension of this idea.

\begin{proposition}[Principle of Ontological Independence]
Any ontological reality in the physical world is independent of the way it is described. In other words, physical reality is invariant with respect to its description. In symbols, if $\mathcal{M}$ and $\mathcal{M}'$ are valid descriptions (models) of $\textbf{E}$, then any true proposition in $\mathcal{M}$ is not false in $\mathcal{M}'$. Formally:
\begin{equation}
 p \in \mathcal{M} \Longrightarrow \neg p \notin \mathcal{M}'
\end{equation}
where $p$ is a proposition, and $\neg p$ is the negation of $p$.
\label{prop:independenciaOnto}
\end{proposition}

This principle is logically equivalent to the following:

\begin{corollary}
The set $\textbf{E}$ possesses a collection of ontological rules, in other words, it has an ontology.
\end{corollary}

Indeed, if $\textbf{E}$ didn't have its own structure, any model endowed with logical consistency would be true in $\textbf{E}$.

Similarly, if there were no determined empirical data, any model $\mathcal{M}$ endowed with logical consistency would trivially be true in $\textbf{S}$. However, because concrete empirical data exist, not just any model will be valid.


\section{Exemplification of the Formalism}

A good example of two physical theories where this principle forces us to discard one of them is the case of gravity according to Newton and the general theory of relativity.
The Newtonian model is as accurate as the general theory of relativity within a certain range of precision and for velocities much less than the speed of light. However, this only refers to the validity of the mathematical description $(\mathcal{M}, I)$. It's not the case regarding the ontological description $(\mathcal{M},O)$ of the respective theories, as one of them postulates that gravitational attraction occurs due to action at a distance -gravity is a force- while the other explains this attraction through the deformation of spacetime -gravity is only an apparent force-.
Thus:
\begin{equation}
    (\mathcal{M},I)_{\text{Newton}} \approx (\mathcal{M},I)_{\text{Einstein}}
\end{equation}
for the regime of small velocities, but
\begin{equation}
    (\mathcal{M},O)_{\text{Newton}} \neq (\mathcal{M},O)_{\text{Einstein}}
\end{equation}
in any regime.
All this implies that the physical interpretation is different according to Newton compared to Einstein:
\begin{equation}
    F_{\text{Newton}} \neq F_{\text{Einstein}}.
\end{equation}
Therefore, if the curvature of spacetime as the cause of gravitation or Newtonian action at a distance are to be accepted as realities, then according to the principle of ontological invariance, their existence cannot depend on the description we choose, and therefore at least one of the two theories is invalid.

To illustrate these ideas, let's provide a concrete example.

Let $\mathcal{M}$ be a structure constructed as follows:

\begin{enumerate}
    \item The objects of $\mathcal{M}$ are called \textit{bodies}.
    \item Bodies occupy one and only one point in Euclidean space $\mathbb{R}^3$.
    \item Every body has a quantitative property called $m$.
    \item Any two bodies attract each other with a force $\vec{F}$ given by
    \begin{equation}
        \vec{F} = \alpha\frac{m_1 m_2}{r^2}(\vec{r_1}-\vec{r_2}) 
    \end{equation} 
    where $m_1$ and $m_2$ are the $m$ of the first and second body respectively; $r_1$ is the position of the first body, $r_2$ is the position of the second body, both in $\mathbb{R}^3$, $\alpha$ is a constant determined by measurement units, and $r$ is the distance between the two bodies, according to the Euclidean metric of $\mathbb{R}^3$.
\end{enumerate}

At first glance, we might say that this theory describes the gravitational interaction between bodies in a Newtonian manner. However, for this to be the case, an empirical interpretation $I$ of the model $\mathcal{M}$ is necessary. Let's define this empirical interpretation:

Let $I$ be a relation defined as follows:
\begin{enumerate}
    \item We associate the quantity called $m$ in the model $\mathcal{M}$ with the empirical quantity that determines the acceleration of a sensible object given a force, which is also known as \textit{mass}.
    \item We associate the Euclidean metric of $\mathbb{R}^3$ in the model $\mathcal{M}$ with the lengths measured in our sensible physical environment, in such a way that a distance between two points in $\mathbb{R}^3$ corresponds to an empirically measured distance.
    \item We associate $\vec{F}$ in the model $\mathcal{M}$ with the \textit{force} of attraction perceptible between different bodies with mass, which is known to be gravity, and is attributed to the empirical fact that things fall toward the ground on Earth.
\end{enumerate}

The reader might think that the interpretation $I$ is unnecessary, as it is implicitly contained in the model $\mathcal{M}$. However, that's not the case, and to demonstrate this, let's propose a second interpretation of $\mathcal{M}$, denoted $I'$:

Let $I'$ be a relation defined as follows:
\begin{enumerate}
    \item We associate the quantity called $m$ in the model $\mathcal{M}$ with the empirical quantity that measures the magnitude of the electric charge of an object.
    \item We associate the Euclidean metric of $\mathbb{R}^3$ in the model $\mathcal{M}$ with lengths measured in our sensible physical environment.
    \item We associate $\vec{F}$ in the model $\mathcal{M}$ with the \textit{force} of attraction or repulsion perceptible between different bodies with electric charge, known as the Coulomb force.
\end{enumerate}

As can be seen, the empirical interpretation $I'$ is entirely different from interpretation $I$, yet both share the same model $\mathcal{M}$ and both are valid within their respective domains of experimental precision.

Now, returning our focus to the model $\mathcal{M}$, an ontological interpretation must be defined to establish the relationship of the model with reality.

Let $O$ be the ontological interpretation defined as follows:
\begin{enumerate}
    \item For each \textit{body} in $\mathcal{M}$, there exists a real entity in $\textbf{E}$ called a \textit{real body}.
    \item Real bodies have a location and move in a space that can be sufficiently approximated by $\mathbb{R}^3$.
    \item Every real body has a real and quantitative property corresponding to the property $m$ in the model $\mathcal{M}$.
    \item There exists an attractive force between real bodies, the magnitude of which can be determined with sufficient accuracy by a mathematical relationship.
\end{enumerate}

Note that this particular ontological interpretation is indifferent to which empirical interpretation is used (whether $I$ or $I'$). The ontology tells us to what extent the model $\mathcal{M}$ can be associated with reality. For example, as stated in item 4 of the definition of $\mathcal{M}$, it was claimed that bodies occupy one and only one point in space. This implies their non-extension. However, this assertion is not associated by interpretation $O$ with any reality; therefore, this assertion is left as a mere mathematical simplification. Thus, $O$ allows the model to be valid even when real bodies are extended and occupy volume. If the ontological interpretation had associated the non-extension of the bodies in $\mathcal{M}$ with a real property, the corresponding physical theory would be falsified at that point if the existence of real bodies with non-zero volume were proven, or it would be obliged to deny the name of <<bodies>> to those objects.

The need to affirm the real existence of the abstract objects in $\mathcal{M}$ through an ontology may seem redundant, but it's not. This becomes evident when instead of speaking about \textit{bodies}, we talk about objects less accessible to the senses, such as atoms, elementary particles, or the \textit{strings}, \textit{loops}, and \textit{branes} of string theories. Suppose $\mathcal{M}$ is the model of a string theory $\mathcal{F}$, and $O$ is the ontology that contains this statement:

\begin{itemize}
    \item \textit{For every string in $\mathcal{M}$, there exists a real object in $\textbf{E}$ corresponding}. (Meaning that strings are real objects).\\
\end{itemize}

Now, let $O'$ be an ontology for $\mathcal{M}$ that doesn't contain such a statement. For $O$, the strings of the string theory model are real entities, existing outside the mind and fundamental elements in the fabric of the universe. However, for $O'$, strings are abstract objects, not necessarily real, which can be used to successfully describe physical phenomena. The difference in perspective between $O$ and $O'$ has consequences even at the experimental level. According to $O$, there might come a day when such strings are \textit{observed}, as they would be talking about real entities, while for $O'$ such a search wouldn't make sense. The current problem of quark confinement is a similar case to our example. Are quarks real entities as realists assume (ontology $O$), or are they not, as instrumentalists assume (ontology $O'$)? Similar problems extend to other particles and fields.

In conclusion, the relationship between the ontological and empirical interpretations is intricate. The existence of objects beyond the empirical realm is always an ontological problem, as empirical data never directly refer to existence. Existence, in itself, cannot be sensed or measured but can only be inferred from sensations and measurements\footnote{The fact that the existence of a real entity is not directly perceived by the senses, but is inferred from them, has been noted by many philosophers since antiquity. Both Aristotle and medieval scholastics argued that it is not possible to directly attain knowledge of substances, but rather indirectly through the perception of their sensory qualities. Descartes, on his part, attempted to establish his philosophical system by deducing his own existence, inferring it from the perception of thought: \textit{I think, therefore I am}. However, he failed in his attempt to demonstrate the existence of the physical world without resorting to inference based on sensory perception.}. Due to this, an empirical interpretation $I$ cannot contain statements that relate the objects of $\mathcal{M}$ to real entities, but only to the phenomenological entities of $\textbf{S}$. 

Similarly, the empirical interpretation cannot include statements about the universality of properties unless that universality is sensorially accessible. For that to happen, all possible cases must be empirically verified. If not, the generalization of a property falls into the realm of ontology. For example, the interpretation $I$ in our first example relates the mass of the bodies in $\mathcal{M}$ to a property that is measured in a certain way. However, there is no empirical guarantee that every body must have mass since only a fraction of them are within the reach of our experimental verification. The generalization that mass is a universal property present in any particle of fermionic matter is an ontological presupposition.

\chapter{Ontology of Change, Space, and Time}
\section{Ontological Reality of Change}
Things change, there is no doubt about that. Denying the ontological reality of change would be equivalent to asserting that all the movements and transformations we observe in nature are mere illusions. This would immerse us fully into Parmenides' monism and distance us from the underlying perspective of modern science. However, before establishing the principle that change is an ontological reality, we must clarify the meaning of such an assertion. Change is not an entity but a fact; it is not something that exists but something that happens. Because of this, we cannot say, <<Let $x \in \textbf{E}$ be a change>>, such a formula makes no sense, as $\textbf{E}$ has been defined as the set of real \textit{beings}, but \textit{facts} are not beings. A fact doesn't correspond to existence but to occurrence; facts don't \textit{are}, they \textit{happen}. 

With this in mind, we will model change as follows. Let $x \in \textbf{E}$ and suppose a real change occurs in which $x$ transforms into $y$. We will then write
\begin{equation}
x \longrightarrow y.
\label{eq:note}
\end{equation}
Equivalently, we can say that in change, a given entity $x$ moves from a state $a$ to a state $b$:
\begin{equation}
x_a \longrightarrow x_b.
\label{eq:note2}
\end{equation}
Here, $x_a$ represents entity $x$ in state $a$, and similarly, $x_b$ represents entity $x$ in state $b$. $a$ and $b$ are elements of some set of possible states. Entities capable of undergoing changes will be termed \textbf{mutable}.

Now, let's consider that the mutable entity $x$ undergoes a series of changes, not necessarily discrete:
\begin{equation}
x_a \longrightarrow x_b \longrightarrow x_c \longrightarrow \cdots
\end{equation}
with $a, b, c, \ldots$ as elements of a set of possible states $X$. We maintain that the following proposition is true.

\begin{proposition}(Transitivity of Change)
Let $a$, $b$, and $c$ be possible states for an entity $x$. If $x_a \longrightarrow x_b$ and $x_b \longrightarrow x_c$, then it follows that $x_a \longrightarrow x_c$.
\end{proposition}

If we consider the different possible states in a sequence of changes as \textbf{\textit{locations}}, then the elements of the set $X$ will be called \textbf{\textit{points}}, and the set $X$ itself will be referred to as the \textbf{\textit{state space}}.

\section{Motion}
Let $x$ be a mutable entity, and let $a$ and $b$ be possible states for $x$. When $x_a$ is currently present, one of the following two statements can be true:
\begin{enumerate}
    \item It is possible that $x_a \longrightarrow x_b$.
    \item It is happening that $x_a \longrightarrow x_b$.
\end{enumerate}
In case (1), the change to $x_b$ is not determined, as many other changes could also be possible at the same time. In case (2), the change to $x_b$ is \textit{actually} occurring, meaning that the determination of an ongoing change towards $x_b$ is already contained in $x_a$.
We will refer to this determination that distinguishes between mere possibility and the fact of an ongoing change as \textit{enérgeia}\footnote{Although this enérgeia linked to motion could be identified with \textit{kinetic energy}, we will refrain from equating it with any known concept in physics for now. We will examine its properties later on.}. The change that allows this type of determination will be termed \textbf{motion}, and the absence of motion will be termed \textbf{rest}. Entities capable of undergoing motion will be called \textbf{movable} entities. The states that can be reached through motion are referred to as \textbf{locations}, and the set of all possible locations is called the \textbf{space}.

If $x$ currently occupies a location at some point in space $X$, we will say that $x$ is \textit{in} $X$. However, it's important not to confuse the notion of being in a space with the notion of being a point in space. If the mutable entity $x$ is in space $X$, it doesn't mean that $x$ is a point of $X$. We will usually denote the set of entities from $\textbf{E}$ that are currently in space $X$ as $\textbf{E}|_X$. Keep in mind that $\textbf{E}|_X$ is a subset of $\textbf{E}$, not of $X$. Suppose an entity $x$ occupies one, and only one point in space $X$, then we say that $x$ is a \textbf{point-like particle} in $X$.

We will represent motion using the same notation as expressions (\ref{eq:note}) and (\ref{eq:note2}). It's important to emphasize that this notation doesn't represent a mathematical transformation but rather an ontological transformation (i.e., a real change in entity $x$). This means that it must satisfy the principle of ontological independence established in proposition \ref{prop:independenciaOnto}. Now, applied to the particular case of motion, this principle is stated as follows:

\begin{proposition}
Motion is an ontological reality. Therefore, the distinction between rest and motion is an ontological invariant.
\label{principio:movimiento}
\end{proposition}

However, it should be noted that while motion is an ontological reality, it doesn't prevent it from being a relative reality. A relative property is always stated in relation to something else, different from the subject. Thus, it's entirely plausible for a passenger to be at rest relative to the airplane they are traveling in, yet be in motion relative to the ground. This doesn't contradict proposition \ref{principio:movimiento} because
\begin{eqnarray}
p&:=&\text{'the passenger is moving relative to the ground', and} \\
q&:=&\text{'the passenger is not moving relative to the airplane'}
\end{eqnarray}
are two distinct, non-equivalent statements, and therefore, the ontological principle from proposition \ref{prop:independenciaOnto} is not violated. According to our ontology, if a valid model asserts that an object is in motion, it cannot be contradicted by any other valid model. However, for this to hold, the statement 'the object is in motion' needs to be formulated in an absolute manner (i.e., independently of the choice of reference frame).

\subsection{Fundamental Measures of Change}

Let's examine the possibility of defining a distance function $d$ between any two points in space $X$. If this is possible, then $X$ is a metric space, and the distance between point $a$ and point $b$ will be denoted as $d(a,b)$.

In a change like $x_a \longrightarrow x_b$, there can be a series of intermediate changes through which $x$ passes while moving from state $a$ to state $b$. If $S_{a,b}$ is the set of all those states, we say that $S_{a,b}$ is the \textbf{trajectory} of $x$ in the change $x_a \longrightarrow x_b$.

\begin{definition}
Let $S = S_{a,b}$ be the trajectory of $x$ in a change $x_a \longrightarrow x_b$, and suppose that $S$ is a measurable set. Then, the measure $\mu$ defined over this set is called the \textbf{displacement}. The displacement between states $a$ and $b$ is denoted as $s(a,b)$ and is given by
\begin{equation}
    s(a,b) = \mu(S).
\end{equation}
\end{definition}

\noindent When states $a$ and $b$ and the measure $\mu$ are understood from the context, we will write $\Delta s$ instead of $s(a,b)$. Similarly, when state $a$ is considered as the initial state of the change process, we will write $s$ instead of $\Delta s$.

Depending on the topology of space $X$, an entity $x \in \textbf{E}|_X$ might have different possible trajectories in its path from $a$ to $b$. Let $S_{a,b}^* = \{S_{a,b}\}_{a,b\in X}$ be the collection of all possible trajectories between points $a$ and $b$ in space $X$.

\begin{definition}
The \textbf{distance} between $a$ and $b$, denoted as $d(a,b)$, is the magnitude of the minimum possible displacement between these two points. Formally,
\begin{equation}
    d(a,b) = \min\{\mu(S_{a,b}):S_{a,b} \in S_{a,b}^*\}.
    \label{eq:dist}
\end{equation}
\end{definition}

\noindent There are some important observations to make here. The existence of the minimum in an ordered set implies its uniqueness, meaning that there is a unique value for $d(a,b)$. This implies that the function $d:X^2 \longmapsto \mathbb{R}$ is well-defined.

\begin{theorem}
The function $d$ is a metric on space $X$.
\label{teor:espaciometrico}
\end{theorem}

\noindent The proof of this theorem can be found in the appendix.

These definitions also have an interesting ontological sense. It's worth noting that distance has been defined in terms of displacement, not the other way around. This implies the conception that in our model, displacement is a more fundamental quantity than distance. Therefore, distance should be defined in terms of displacement, not the other way around, as is commonly done.

Another important consequence of these definitions is that the metric $d$ must be positive. This excludes from the outset any pseudoriemannian metric, like the Minkowski metric, capable of yielding negative values\footnote{As we will see, this won't be a problem, since for now, we want to model space, not spacetime, as we haven't introduced the notion of time yet.} (or complex values in the case of norms). Allowing negative values in $\mu$ would render $d$ ill-defined.

The idea that these definitions aim to capture is that of distance as a measure of how \textit{different} states $a$ and $b$ are, and this is achieved by measuring the set of minimal changes required to move from state $a$ to state $b$. The more changes are required, the \textit{farther} $b$ is from $a$. Distance refers to what varies, to the difference.

\subsection{The Measurement Problem for Continuous Trajectories}

Let's consider the case of a change $x_a \longrightarrow x_b$ with the trajectory $S = S_{a,b}$. If this trajectory consists of a finite number of elements, in other words, if it's a discrete trajectory, then the measure $\mu$ can be defined such that $\mu(S) = |S|$, where $|S|$ is the number of elements in $S$. In this case, the \textit{size} of the trajectory is intrinsically determined.

A different situation arises when $S$ is a continuous trajectory. In this case, the \textit{size} of $S$ is indeterminate, and we need a \textit{unit of measurement} to measure the length of $S$.

Here, a profound problem emerges. In order to assign that unit of measurement or \textit{rule}, we need to associate the endpoints of the rule with the endpoints of trajectory $S$. The first thing we might think of is placing one endpoint very close to the start of the trajectory, and seeing at which point along the trajectory the other endpoint of the rule ends up being closest. The problem is that this cannot be achieved because in order to measure distances, we must first be able to measure trajectories. So, we are back to the same problem of how to measure the distance between the endpoints of a rule and points on our trajectory. And so on, \textit{ad infinitum}. In this way, it is intrinsically impossible to assign a measure to a continuous trajectory if we want to associate the endpoints of the rule with two points of trajectory $S$ through a proximity relationship. Consequently, in order to utilize a unit of measurement, the endpoints of our rule must be associated with the points of trajectory $S$ through a different relationship that doesn't involve distinguishing distances.

\subsection{Simultaneity}

\begin{figure}[!h]
\centering
\includegraphics[scale=.8]{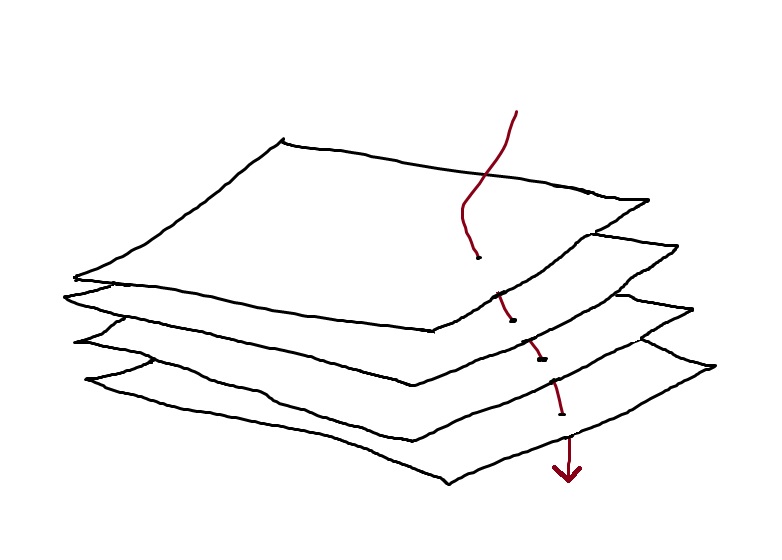}
\caption{\small Representation of the partition induced by the relation $\sim$ as a fibered space. Entities that are in the same fiber are simultaneous. The red line represents the trajectory of a moving entity.}
\label{fig:bundle}
\end{figure}

Let $x$ and $y$ be two entities such that $x_a \longrightarrow x_b$ and $y_i \longrightarrow y_j$. Let $x_u$ be an element of the trajectory of $x$ and $y_v$ be an element of the trajectory of $y$, in their respective changes.
The relation of \textbf{simultaneity} is an equivalence relation between the elements $x_u$ and the elements $y_v$.
We will write $x_u \sim y_v$ to express that $x_u$ is \textbf{simultaneous} with $y_v$.
Like any equivalence relation, simultaneity is reflexive, symmetric, and transitive. Every equivalence relation induces a partition in the set where it is defined. In this case, the relation $\sim$ induces a partition where the space $X$ can be conceived as a fibered space, where each fiber is an element of the quotient $X/\sim$.
Figure (\ref{fig:bundle}) shows a representation. If $x_a$ and $y_b$ are in the same fiber, then $x_a \sim y_b$ holds; if they are in different fibers, then $x_a \not\sim y_b$. The fibers of simultaneity do not have contact points and are disjoint from each other.

This concept allows us to introduce the notions of duration and time. For this, we need a reference change; the object that undergoes such a change will be called a \textbf{clock}.

\begin{definition}
Let $x$ be an entity that undergoes the change $x_a \longrightarrow x_b$. Let $r$ be a clock such that $r_i \longrightarrow r_j$, where $x_a \sim r_i$ and $x_b \sim r_j$.
We call the \textbf{duration} of the change $x_a \longrightarrow x_b$ according to the clock $r$, the trajectory $S_{i,j}$ of $r$.
\end{definition}

\begin{definition}
Let $D_r$ be the duration of a change $x_a \longrightarrow x_b$ according to the clock $r$. We call \textbf{time} the measure $\mu(D_r)$, i.e., the displacement of the clock
\begin{equation}
 \mu\left(D_r\right) = s_r(i,j) \qquad  x_a \sim r_i, x_b \sim r_j.  \end{equation}
\end{definition}
We will write $t(a,b)$ for simplicity, which is defined as
\begin{equation}
    t(a,b) = \mu(D_r).
\end{equation} 
When the points $a$ and $b$ and the measure $\mu$ are understood, we will write $\Delta t$ instead of $t(a,b)$. Similarly, when the point $a$ is considered the initial state of the change, we will write $t$ instead of $\Delta t$.

\begin{figure}[!h]
    \centering
    \includegraphics[scale=.5]{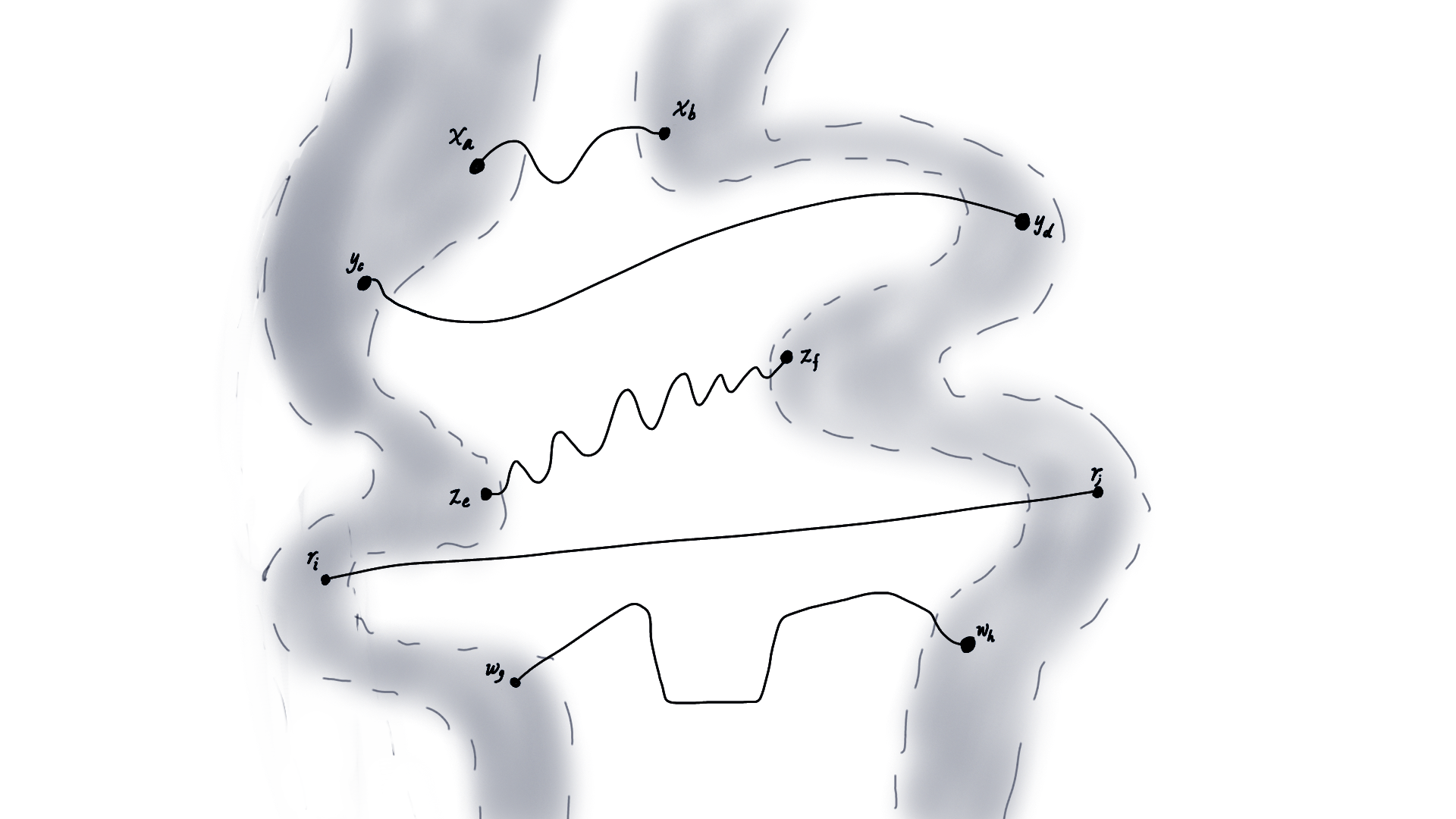}
    \caption{\small In this representation, shaded areas represent fibers of simultaneity. Lines represent trajectories, and points represent the endpoints of a change. If we take the entity $r$ as a clock, we can see that all represented changes have the same duration, even though they have different trajectories.}
    \label{fig:diagr_1}
\end{figure}

\begin{definition}
Let $x$ and $r$ be elements of $\textbf{E}|_X$. Let's consider that $r$ is a clock for $x$, and that $x_u \sim r_v$, where $u$ and $v$ belong to the trajectories of $x$ and $r$ respectively. We say that $r$ is the **proper clock** of $x$ if $d(u,v) \approx 0$. In this case, the time $t(a,b)$ is called the **proper time** of $x$.
\end{definition}
In simple terms, the proper time of $x$ is the time measured by a clock traveling arbitrarily \textbf{close} to the entity $x$.

It is not necessary for a clock to move in the same space as the entity that uses it as a reference. A change $x_a \longrightarrow x_b$ with $a,b\in X$ can measure its duration using a clock $r$ such that $r_i \longrightarrow r_j$ with $i,j \in Y$, where $Y \neq X$. The only requirement is that it is possible to establish the simultaneity relationship $x_a \sim r_i$ and $x_b \sim r_j$, as well as all their intermediate states. On the other hand, it may be the case that $r_g \longrightarrow r_h$ with $g,h \in X$, without this change being the reference for measuring the duration.

\subsection{Illustration of Concepts}

Let's consider an example. Suppose we have a spaceship $x$ moving in space, and let $r$ be a chronometer. The change of the spaceship involves moving from a point $a$ to a point $b$ in the space $X$. In this example, let's interpret $X$ as interplanetary space, modeled by $\mathbb{R}^3$.

The chronometer $r$ is on board the spaceship, so we can say that it measures the \textit{proper time}. Let $Y$ be the space of all values that this chronometer can take. Let's assume it measures time with an accuracy of tenths of a second (36000 tenths of a second in total). Then, the space where the changes of the chronometer take place can be modeled by a sequence of values from 0 to 36000. $Y = \{0,1,...,36000\}$.

The trajectory $S_r$ of $r$ in $Y$ will be the subset from value $n$ to value $m$, and the displacement in $Y$ will be the number of chronometer steps taken in this trajectory, i.e., $s(n,m)=m-n$.

Suppose our spaceship takes off from Earth, goes into orbit, and we want to measure the time it takes for this change in location. If $x_a$ represents the spaceship at the moment of liftoff, and $x_b$ represents it at the moment of \textit{burnout}, then the duration of going into orbit $x_a \longrightarrow x_b$ according to the chronometer $r$ is the displacement of the chronometer in the space $Y$. In this case, liftoff is simultaneous with the start of the chronometer $x_a \sim r_0$, while burnout (when the engines are turned off) is simultaneous with stopping the chronometer $x_b \sim r_j$. Let's assume that at the moment of burnout $j=1800$, then the duration of the launch in this model is $t(a,b)=s(0,1800) = 1800$ tenths of a second.

The reference change used to measure the duration of the launch takes place in the space $Y$, but nothing prevents the chronometer $r$ from moving (changing location) in the space $X$ at the same time, where it coexists with the spaceship. However, the change $r_g \longrightarrow r_h$ where $g$ and $h$ are points in $X$ was not the reference for measuring the duration of the launch. If it had been, the chronometer's displacement would have coincided with the spaceship's displacement within which it was located.

It might seem that an entity cannot move in the same space and be useful at the same time in that space as a clock for measuring the duration of some change. We will show that this is false. Let $x$ be the same spaceship as in the previous example, and let $y$ be a beam of light coming from the Sun that passes near Earth. Let's choose $y$ as our clock. So when the spaceship takes off $x_a$, the beam of light is at a point $g$ in space, so $x_a \sim y_g$. Then, at the moment of burnout, the spaceship is at point $b$, represented by $x_b$, and the beam of light will be at point $h$ in the same space, represented by $y_h$. We have $x_b \sim y_h$. Therefore, the duration of going into orbit, using the beam of light $y$ as a clock, is the displacement $s(g,h)$. In this case, assuming that light follows the shortest path, the displacement is given by the distance between these two points, let's say $d(g,h) = 53820 \text{ km}$.
Thus, with respect to this clock, the duration of the launch will be $t(a,b) = 53820 \text{ km-Light}$. Note that just because the beam of light was chosen as the clock, its displacement, which is given in units of distance, should be considered as a unit of time: $53820 \text{ km-Light} \approx 1800 \text{ tenths of a second}$.

\section{The \textit{Paradox} of Simultaneity, the Rule, and the Clock}

As previously explained, the essential difference between a clock and a ruler is that when measuring with a ruler, we obtain the displacement of the measured object, while when measuring with a clock, we obtain the displacement of the clock itself. It would seem natural for the same entity to function as both a ruler and a clock simultaneously. However, this implies that we are capable of knowing the fibers of simultaneity prior to any measurement of displacement or duration.

Unfortunately, this is impossible, as empirically perceiving ontological simultaneity is beyond our reach. To know it, we need measurements of length and time, and thus rulers and clocks. In this case, the same entity cannot function as both a ruler and a clock simultaneously, as, without knowing simultaneity, to define a unit of time, we would need a ruler (distinct from the clock). Similarly, to define the unit of distance, we would need a clock (distinct from the ruler).

\begin{proposition}
Let $x_a \longrightarrow x_b$ and $y_{a'} \longrightarrow y_{b'}$ be changes, with $a, a', b, b' \in X$. Suppose that we have no way of determining if $x_\alpha \sim y_{\alpha'}$ for any state $\alpha\in X$ except through measurements. Then, it is not possible to use $y$ as both a ruler and a clock simultaneously to measure the displacement of $x_a \longrightarrow x_b$ and its duration.
\end{proposition}

A concrete example will make this difficulty clearer. Suppose that $x$ is the spacecraft from the previous section, and $y$ is light, which we wish to use as both a ruler and a clock simultaneously. Recall that we are incapable of determining the ontological simultaneity of two events except through measurements (for which we need rulers and clocks). Let's first try to define our unit of distance, which we will call the \textit{meter}. The official definition of the meter works perfectly here:

\begin{quote}
The meter is the length of the path traveled by light in a vacuum during a time interval of 1/299,792,458 of a second \cite{sistema2006}.
\end{quote}

In other words, a meter is 1/299,792,458 light-seconds.

The problem arises when we try to use the same entity (in this case, light) to define our unit of time (the second). To do that, we would need to define it as the time it takes for light to travel from point $a$ to point $b$. However, since ontological simultaneity is inaccessible without measurement, we would need a ruler to measure the distance between these two points and then define the second as follows:

\begin{quote}
The second is the time interval of the duration of the path traveled by light in a vacuum over a distance of 299,792,458 meters.
\end{quote}

Clearly, we would have fallen into a circular definition, defining the meter in terms of the second and the second in terms of the meter. As can be seen, we can use light as a ruler, but in such a case, we need to find a distinct clock. This is exactly what is done in the International System of Units (SI), where light is used as a ruler but not as a clock. Instead, the second is defined based on the oscillations of the cesium atom\footnote{According to the SI, the second is the duration of 9,192,631,770 periods of the radiation corresponding to the transition between the two hyperfine levels of the ground state of the cesium 133 atom \cite{sistema2006}.}.

It would also be possible to use light as a clock, defining the second as the duration of a specific path of light, as done earlier. However, in this case, we would need a distinct entity to define the meter.

\begin{figure}[!ht]
    \centering
\begin{displaymath}
    \xymatrix{--&&y_{b'}\ar[drr]^\sim&&\\
		        &--&y_b \ar[u]\ar[rr]^{\text{COES}}& & x_b \ar[ull] \ar[ll]\\
         &\perp\ar[u]^{\text{dist.}}&y_a\ar[rr]^{\text{COES}}\ar[u]&& x_a\ar[u]\ar[dll]^\sim\ar[ll]\\
         \perp\ar[uuu]^{\text{duration}}&&y_{a'}\ar[urr]\ar[u]&&}
\end{displaymath}
    \caption{Representation of an entity that is both a ruler and a clock, and how it determines distance and duration of a change. The simultaneity relation is marked with $\sim$, while $\text{COES}$ indicates the relation of entities that are co-spatial (sharing the same point in space, not necessarily simultaneously). The distance from $a$ to $b$ is the displacement from $y_a$ to $y_b$, whereas the duration of the change $x_a \longrightarrow x_b$ is the displacement from $y_{a'} \longrightarrow y_{b'}$.}
    \label{fig:relojregla}
\end{figure}
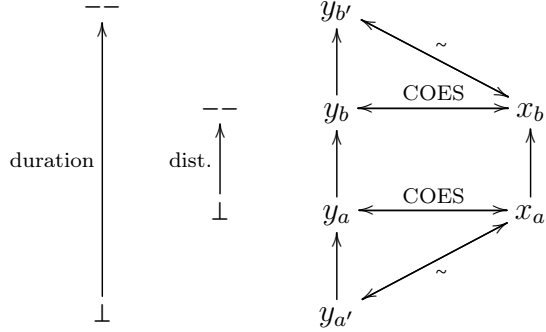

This problem would disappear if we were able to determine ontological simultaneity without measurement. Then, for a change $x_a \longrightarrow x_b$, the distance would be the displacement of light traveling from point $a$ to point $b$, and the duration of the change would be the displacement of light traveling from a point simultaneous with $x_a$ to a point simultaneous with $x_b$. Figure \ref{fig:relojregla} shows a diagram of these relationships.

\section{Consequences}
The definition of time as a measure of the displacement of a clock can be perplexing; however, it is appropriate. This conception of time allows us to describe the most basic concepts of physics without separating them from their ontological content.
Thus, we can measure the movement of a mobile entity $x$ as the difference or \textit{gradient} between the displacement of $x$ and the clock $r$. We will call this gradient \textbf{velocity}.
Mathematically, the gradient describes the rate of change of one variable with respect to another. There are different ways to define it, depending on the mathematical model we are using to describe space. We will usually use the notation $\frac{\partial u}{\partial v}$ to refer in an abstract way to any gradient between $u$ and $v$. Thus, in general, velocity would be defined as
\begin{equation}
     v := \frac{\partial s}{\partial t}
\end{equation}
where $s$ is displacement and $t$ is time.
An appropriate definition in this case would be through the ordinary derivative
\begin{equation}
    v = \frac{\partial s}{\partial t} =\frac{ds}{dt}.
\end{equation}

Now let's consider an entity $y$ that functions as both a clock and a ruler. Then, the velocity of this entity when it moves from point $a$ to point $b$ will be
\begin{equation}
v = \frac{ds(a,b)}{dt(a,b)} = \frac{ds(a,b)}{ds(a,b)} = 1.
\end{equation}

\noindent This is because time is equal to the displacement of the clock, $t(a,b) = s(a,b)$.
It follows that such an entity should have a constant velocity.

\chapter{Space}
In the previous chapter, we made use of the idea that two mutable entities $x$ and $y$ can be at the same point $a$ in space $X$, but not simultaneously. This idea was essential in the structure of our model. It would be impossible to determine the distance between two points in space if there were no way to relate the ends of our ruler to such points, in the same way that it is impossible to determine the duration between the ends of a change if we cannot establish a relationship of simultaneity between the clock and the mentioned ends.

In the same manner that the existence of simultaneity fibers is required, the existence of co-space fibers is required too. These fibers would be nothing but the same points of space $X$ persisting in time, either as mere possible locations or as the actual location of some mutable entity.

\begin{figure}[!h]
    \centering
    \includegraphics[scale=.4]{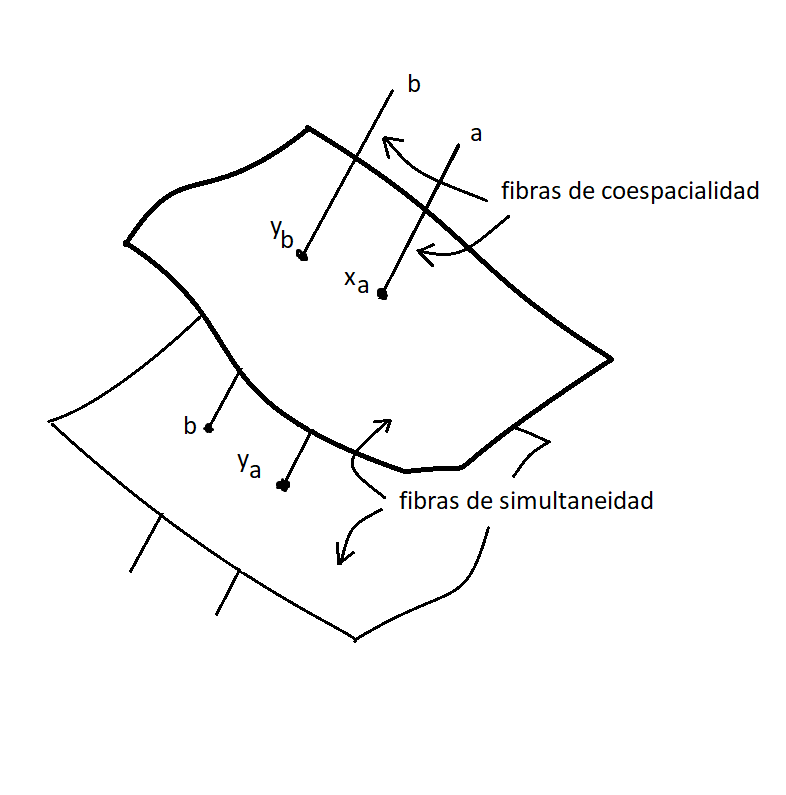}
    \caption{Each point in space has a co-space fiber that associates that point with a simultaneity fiber.}
    \label{fig:coespacialiad}
\end{figure}

A co-space fiber $f$ can be conceived as a curve whose points represent a unique point $a$ in space $X$, associated with a unique simultaneity fiber. Note that the points of $X$ cannot persist in time since the simultaneity fibers are disjoint. See Figure \ref{fig:coespacialiad}.
In this way, time is measured between simultaneity fibers, and distance between co-space fibers.

According to this, co-space fibers would be like the <<trajectories>> of the points in space $X$ through time. Note that the points of $X$ are not mobile entities.

From an ontological standpoint, the concept of a fibered space is not relevant; rather, the concept of co-spatiality is significant, which simply means that two entities are at the same point in space (not necessarily simultaneously). In Figure \ref{fig:coespacialiad}, $x_a$ is co-space with $y_a$, even though they are not simultaneous; on the other hand, $y_b$ is simultaneous with $x_a$ but not co-spatial.

For this to be possible, it is necessary for each point in space to be distinguishable individually, regardless of the moving entity present in it. In this chapter, we will analyze the properties that this implies in the concept of space.

\section{Types of Space According to the Nature of Their Points}\bigskip

\begin{definition}
We say that a space $X$ is \textbf{absolute} when the location of any given particle $x$ within that space does not depend on the location of the rest of the particles inhabiting it. Otherwise, we say that space $X$ is \textbf{relative}.
\end{definition}\medskip

\begin{figure}[!h]
\centering
\includegraphics[scale=.7]{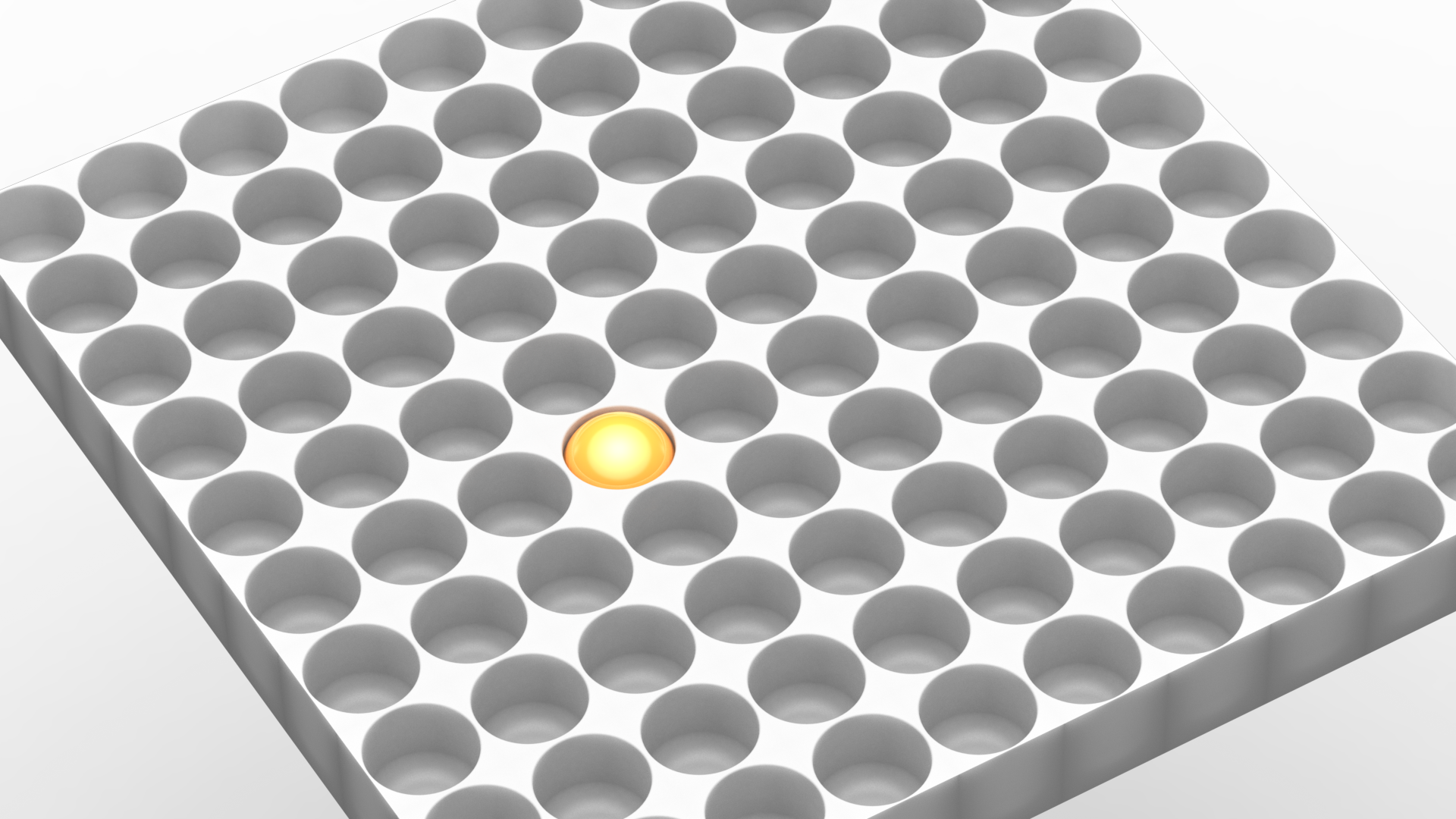}
\caption{\small Example of an absolute and extrinsic space. The holes on the board are the \textit{points} of this space, while the sphere is the \textit{particle}. The only way to establish the location of the sphere is to know its relationship with the hole in which it is placed. Note that the sphere and the hole on the board are distinct entities and exist independently.}
\label{fig:espacioextrinseco}
\end{figure}

An absolute space $X$ is \textbf{extrinsic} if the particles within it do not inherently contain the information needed to establish their location in space. Therefore, this information can only be determined based on the relationship between the particle and the point in space it occupies.
In such cases, the points of $X$ are real entities distinct from the entities existing within that space. If $X$ is a space of this type, then $X \subset \textbf{E}$. Furthermore, if $z \in \textbf{E}|_X$, then there exists a relation $\phi$ that associates the particle $z$ with the point $p$ within $X$ where it is located. Thus, $\phi(z) = p$, for some $p \in X$. If $z$ is at $p$, we say that there is a \textit{coupling} between $z$ and $p$. In an extrinsic space, location is an inherent property of the points in space, and particles only possess location due to their coupling with points. In terms of extrinsic space, it is improper to say that $z$ and $p$ are in the same place because $p$ is not in a place; rather, it \textit{is} the place. An example of this type of space is illustrated in Figure \ref{fig:espacioextrinseco}. In this type of space, the location of a particle depends on its relationship with the space itself and the location of other particles. Particularly, the location of the sphere is determined by its relation with the board and not by the potential locations of other spheres within it.

\begin{proposition}
In an absolute and extrinsic space $X$, change is impossible unless the points of $X$ are mobile entities in another space that is not absolute and extrinsic.
\label{prop:absolutoextrinseco}
\end{proposition}\medskip

\begin{proof}
Let $X$ be an absolute and extrinsic space, and let $x\in\textbf{E}|_X$ be a mutable entity. Suppose that the change $x_a \longrightarrow x_b$ occurs. Then $x_a\not\sim x_b$. However, since $a \in \textbf{E}$ (as $X \subset \textbf{E}$), we must have $a \sim x_a$. Because point $a$ continues to exist when we have $x_b$, we have $a\sim x_b$. Therefore,
\[a \sim x_a \not\sim x_b \sim a\]
which is a contradiction.
\end{proof}

The reason why a physical system like the one in Figure \ref{fig:espacioextrinseco} is possible is that this system is embedded in another space, in which the spheres and also the space itself (the board) move. This space where the board can move is not absolute and extrinsic.

\begin{figure}
    \centering
    \includegraphics[scale=.4]{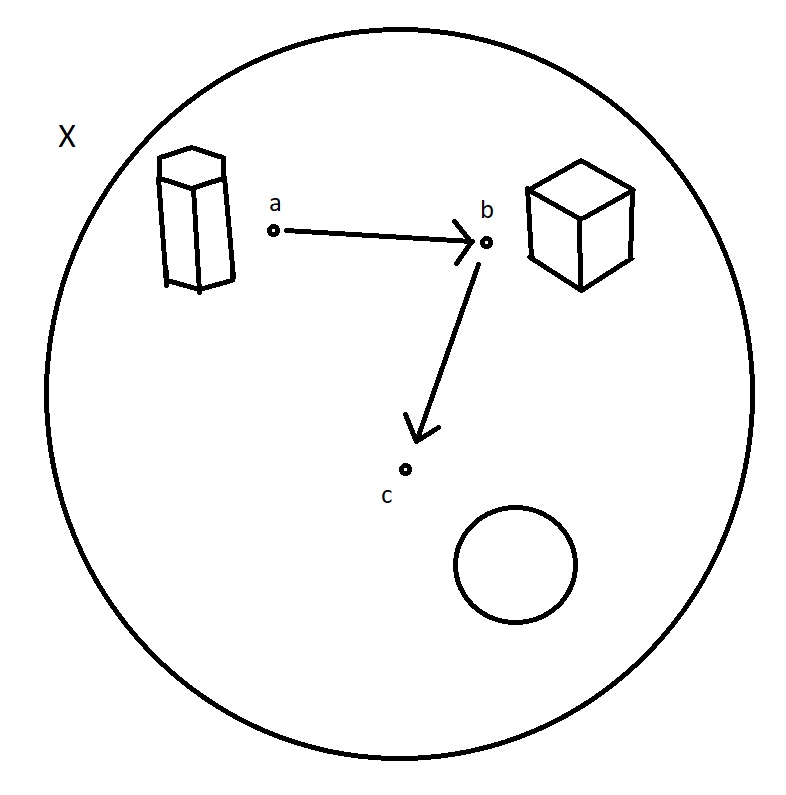}
    \caption{Different points in $X$ represent different shapes that $x$ can take, a trajectory in this space signifies a gradual change of shape.}
    \label{fig:espacioformas}
\end{figure}

An absolute space is \textbf{intrinsic} when it is not extrinsic. In other words, when the particles within it inherently contain the necessary information to establish their location within that space. An example of such a space is the set of possible shapes that a piece of malleable material could adopt (see Figure \ref{fig:espacioformas}). In this case, each possible shape is a point in this space. Note that the shape of the piece of material does not depend on the shape that other pieces might have, which makes the space absolute. Additionally, the piece of material contains its own shape and, therefore, the necessary information to determine its location within the space. Hence, it is an intrinsic absolute space. In intrinsic space, unlike extrinsic space, the point and the particle are not entities that can exist independently of each other.

In a relative space, the location of a particle depends on the locations of others. Therefore, the information to determine particle locations in a relative space is contained within the entirety of those particles. Consequently, a relative space is always intrinsic.

In summary:
\begin{equation}
    \text{Space}\quad\left\{\begin{array}{l}
        \text{Absolute}\quad\left\{ \begin{array}{l}
             \text{Extrinsic}\\
             \text{Intrinsic} 
        \end{array}\right. \\ \\
        \text{Relative}\quad\left\{\text{Intrinsic}\right.
    \end{array}\right.
\end{equation}

\section{Principles of Ontological Invariance}

However, in order to distinguish between different states of motion of a moving entity, a description of such motion is necessary, as well as a way to measure it. Remember that we have referred to the operator used to measure motion as the \textit{gradient}, and we will represent it using the symbol
\begin{equation}
    \frac{\partial u}{\partial v}.
    \label{partial}
\end{equation}
The purpose of this notation is to denote the measure of the rate of change of $u$ with respect to $v$. This \textit{gradient} has to be defined in each model; therefore, the symbol $\partial$ used in the expression (\ref{partial}) does not necessarily represent partial differentiation. The way to measure a gradient of change can vary depending on the mathematical structure of the model, just as the metric can vary depending on the geometry.
The expressions in which we use this notation refer to any way of measuring a gradient, which must be determined by the model in each case, whether it is conventional differentiation, covariant derivative, geometric derivative, etc.\\

\renewcommand{\vec}[1]{\mathbf{#1}}

Let's first analyze the consequences of proposition \ref{principio:movimiento} in an absolute and intrinsic space.
Since the space is intrinsic, we know that the information about the location of a particle is contained within the particle itself. For example, in the case of the malleable material example, we can define a description of the material's shape, let's call it $f$, and a function $u$ that assigns a point in $\mathbb{R}^2$ to each distinct shape $f$. Then, the position $\vec{x}$ of a particle $x$ with shape $f$ would be given by $\vec{x} = u(f)$.\\

\begin{figure}[!h]
    \centering
    \includegraphics[scale=.4]{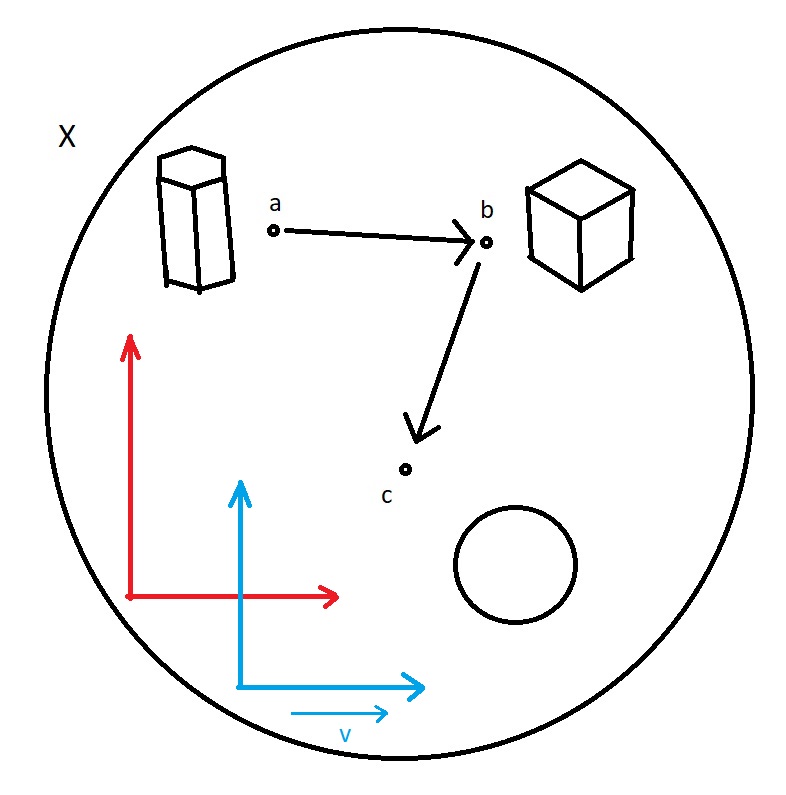}
    \caption{Two different reference frames for determining the positions of points in $X$, one static (in red) and another moving at a constant velocity v (in blue).}
    \label{fig:espacioformascor}
\end{figure}

However, the particle only moves in this space when it changes shape, and this condition must be satisfied for any way of describing the shape, or any function $u$. Thus, if we use a different function to assign a point in $\mathbb{R}^2$ to $f$, let's call it $\bar{u}$, then we will have a new position for the particle: $\bar{\vec{x}} = \bar{{u}}(f)$. 
Nevertheless, the particle will only actually change when it changes shape, leading us to the following expression
\begin{equation}
\frac{\partial\vec{x}}{\partial t} = 0\quad \Longleftrightarrow\quad \frac{\partial \bar{\vec{x}}}{\partial t} = 0.
\label{invariancia:intrinseca}
\end{equation}
Among other things, the expression (\ref{invariancia:intrinseca}) tells us that for ontological reasons, only certain types of functions $u$ should be admitted. In Figure \ref{fig:espacioformascor}, we have illustrated the case of two coordinate assignment functions that do not satisfy the condition (\ref{invariancia:intrinseca}). The first of them, $u_1$, represents a fixed reference frame (in red) with respect to the description of the object's shape, while $u_2$ represents a reference frame moving at a constant velocity (in blue). \\

It can be seen that in a model using $u_2$, objects can change shape without changing position, as they will be fixed with respect to the reference frame moving at a constant velocity. If the intention of such a model is to represent the space of possible shapes, the violation of condition (\ref{invariancia:intrinseca}) highlights an ontological inconsistency.\\

Ignoring this condition would lead us to claim that the reality of the change 
\begin{equation}
    x_{f_1} \longrightarrow x_{f_2}
\end{equation}
for the particle $x$ changing from one shape $f_1$ to another shape $f_2$ ($f_1 \neq f_2$), depends on our description, in direct contradiction with the principle of ontological independence (proposition \ref{prop:independenciaOnto}) and with our assumptions that it is an absolute and intrinsic space.\\

Nevertheless, the relationship (\ref{invariancia:intrinseca}) does not tell us which of the two functions, $u_1$ or $u_2$, should be preferred.
To eliminate this indeterminacy, a second condition is required that establishes the existence of an immobile reference with respect to which we can define when there is movement and when there is not. This object will be referred to as the \textit{natural referent}.
What we need, then, is a point in the space of shapes $X$ that is immobile in itself. Because we are assuming that the space is absolute and intrinsic, any point in this space possesses the property of immutability; a particle can change shape, but a shape will always be that same shape. Thus, the condition that any function $u$ must satisfy can be expressed as
\begin{equation}
\frac{\partial \vec{x}}{\partial t} = 0\quad \Longleftrightarrow\quad \frac{\partial \vec{x}}{\partial u(f)} = 0.
\label{referente:naturalformas}
\end{equation}
where $\displaystyle\frac{\partial \vec{x}}{\partial u(f)}$ expresses the change in the particle's position $\vec{x}=u(x)$ with respect to the position of the shape $u(f)$.\\

In this particular example, the references for distinguishing movement lie within the shapes themselves, and any ontologically consistent model must satisfy the relationships (\ref{referente:naturalformas}) and (\ref{invariancia:intrinseca}). The same concepts logically extend to absolute extrinsic spaces, in which the natural reference points are the points of the space itself, which have their own existence independent of the existence of particles (cf. fig \ref{fig:espacioextrinseco}). However, as has already been demonstrated, in this type of space, movement is impossible (proposition \ref{prop:absolutoextrinseco}).\\

These relationships cease to be trivial when studied in relative spaces, as we will see next.

\section{Consequences of Ontological Invariance in Relative Spaces}

In a relative space, we cannot conceive of motion or rest in absolute terms for particles that move within it. In a relative space, we do not say that a particle just moves; rather, we say that it moves \textit{relative to} another particle or set of particles. As a consequence, the relative motion of a particle cannot be viewed as an ontological reality, as that would violate the principle of independence (proposition \ref{prop:independenciaOnto}). Indeed, an object in a relative space could be in motion or not, depending on the reference frame of our description.\\

For example, let's imagine a universe whose space is relative and whose most suitable mathematical description is $\mathbb{R}^3$, and suppose that the only particles inhabiting this universe are two electrons, let's call them $e_A$ and $e_B$ respectively. Let's assume that these electrons possess spherical symmetry in all their properties.\\

Now then, let's imagine that the two electrons in this universe move away from each other at a constant velocity.\\

The following statements are true:
\begin{itemize}
    \item If we place the origin at $e_A$, then $e_B$ moves away at a constant velocity, while $e_A$ remains at rest.
    \item If we place the origin at $e_B$, then $e_A$ moves away at a constant velocity, while $e_B$ remains at rest.
    \item If we place the origin at the midpoint between $e_A$ and $e_B$, both electrons move away from each other at a constant velocity.
    \item If we place the origin at the midpoint between $e_A$ and $e_B$, and consider both electrons to be at rest, they will both shrink uniformly.
\end{itemize}\medskip

It is clear that proposition \ref{principio:movimiento} conflicts with these statements if we assume that motion in a relative space is a real change of the particle and not of the total set of particles.
However, the ontological principles remain unviolated if we realize that the motion in this type of space is actually a change in the relationships of the particles with each other. In our example, all the propositions are compatible with a description that defines the metric properties of the relationships between the particles. Thus, for example, if instead of two electrons we had $n$ moving entities in our universe, the real change would consist of the change in the distance relationships between any pair $e$ and $e'$ of entities.
\begin{equation}
    D = \{d(r, r') : e, e' \in \textbf{E}|_X\}.
\end{equation}
Where $r = r(e)$ and $r' = r(e')$ are the points in the space $X$ where the entities $e$ and $e'$ are located respectively, and $\textbf{E}|_X$ is the set of all entities populating the space $X$.
Presumably, any change in the relative position of the particles in this space will be a change in the description $D$, and the absence of relative change between the particles implies that $D$ does not change. 
Thus, the description $D$ would be consistent with the principle of ontological independence. We would have
\begin{equation}
\frac{\partial D}{\partial t} = 0 \quad\Longleftrightarrow\quad \frac{\partial \bar{D}}{\partial \bar{t}} = 0,
\label{invariancia:metricas}
\end{equation}
for any two valid descriptions $D$ and $\bar{D}$.
Unlike in the example of the functions $u$ used in intrinsic absolute spaces, the relationship (\ref{invariancia:metricas}) is not fulfilled for arbitrary metrics, as if we define a time-dependent metric $D_1$ and another metric $D_2$ independent of time, ontological invariance might not hold.\\

However, unlike what we did in the previous section, this time we cannot simply discriminate against inadequate metrics, since here there are no \textit{shapes} $f$ at the points of $X$ to refer to.\\

So then, how can we know from two metrics\footnote{Recall from Chapter 2 that the metric of $X$ depends on the chosen measure $\mu$, which in turn depends on the choice of rules and clocks.} $D_1$ and $D_2$ that do not satisfy the relationship (\ref{invariancia:metricas}), if any of them is ontologically valid?
To answer this, we will use the postulate of the existence of the natural referent again: There exists a distance $D_k$ that serves as an unchanging reference for determining whether another distance $D$ changes or not. The distance $D$ is considered to change if and only if its relationship with $D_k$ varies with time. Formally,
\begin{equation}
\frac{\partial D}{\partial t} = 0 \quad\Longleftrightarrow\quad \frac{\partial D}{\partial D_k} = 0.
\label{refnatural:metricas}
\end{equation}
Let's think a bit about the consequences of these statements. Saying that there exists a distance $D_k$ that serves as an immutable reference for defining the other distances leads us to consider the existence of two entities $e_A$ and $e_B$ whose distance for some reason is preferred as a reference point, and consequently is always constant. It can be shown that the maximum number of such particles, whose nature is to maintain the same distance with each other, is $n + 1$, where $n$ is the number of dimensions of the space $X$. In a 3-dimensional space, the maximum number of particles for a natural referent of this kind would be four, and they would be distributed as shown in the figure
\begin{figure}[!ht]
    \centering
    \includegraphics[scale=.5]{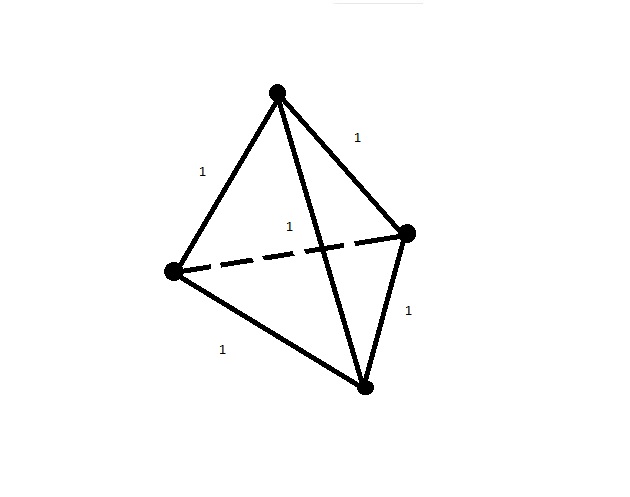}
    \caption{In $\mathbb{R}^3$, 4 equidistant points can be placed.}
    \label{fig:4enR3}
\end{figure}

This way of conceiving the natural referent as if it were an <<Infinity Gem>> presents many problems, both in its ontological interpretation and in its empirical interpretation. Therefore, we will discard it from our study.\\

Let's return to considering the relationship (\ref{invariancia:metricas}). If we are to consider the time dependence, the most general way would be to consider $d$ as a metric in spacetime, which defines a distance between two points in it, that is, between two \textit{events}.
Now, since what we want to know is whether our metric defines a distance whose changes correspond to real changes, independent of the description, we want to know the distance between a particle, let's call it $e$, and that same particle after it has moved through space for a certain time.
The natural referent will then be an entity for which the distance traveled in a unit of time is always invariable, or in other words, it has a constant velocity.
We define \textit{velocity} as
\begin{equation}
    v = \frac{\partial d}{\partial t}
\end{equation}
thus we arrive at the following conclusion\\

\begin{proposition}
Let $v$ be the velocity of a particle in a relative space described by a certain reference frame and with a given metric, and let $\bar{v}$ be the velocity of the same particle in a possibly different reference frame and metric. Then, if both are valid descriptions of reality, the following is satisfied:
\begin{equation}
\frac{\partial v}{\partial t} = 0 \Longleftrightarrow \frac{\partial \bar{v}}{\partial \bar{t}} = 0.
\label{invariancia:relativa}
\end{equation}
\\
And there exists an entity with velocity $k$ such that
\begin{equation}
\frac{\partial v}{\partial t} = 0 \Longleftrightarrow \frac{\partial v}{\partial k} = 0.
\label{referente:natural1}
\end{equation}
\label{prop:invarianciavel}
\end{proposition}

The results contained in proposition \ref{prop:invarianciavel} are not without surprise, as they tell us that only for ontological reasons, in any relative space, there must exist an entity that travels at a constant velocity under any valid description. A conclusion that is by no means trivial from any point of view.\\

In the next chapter, we will use these postulates to test the main physical theories that model space and time.
\chapter{Ontological Test on Various Physical Theories}
\section{Ontological Test on Galilean Transformations}

Let $X$ be a space modeled by Euclidean space $\mathbb{R}^3$. The position of the particles populating space $X$ is described using rectangular coordinates.\\

The Galilean coordinate transformation equations are
\begin{equation}
    \begin{cases}
\bar{x}=  x - V_xt \\
\bar{y}  =  y\\
\bar{z} =  z
    \end{cases}
\end{equation}

Firstly, we define the gradient as the ordinary derivative.
\begin{equation}
    \frac{\partial x}{\partial t} = \frac{dx}{dt}
\end{equation}
We review whether this model satisfies the relationships of proposition \ref{prop:invarianciavel}. The relation (\ref{invariancia:relativa}) remains as
\begin{equation}
\frac{dv}{dt} = 0 \quad\Longleftrightarrow\quad \frac{d\bar{v}}{dt} = 0
\label{invariancia:relativaGal}
\end{equation}
since
\begin{equation}
\frac{d\bar{v}_x}{dt} = \frac{d}{dt}\left(\frac{d}{dt}\left(x - V_xt\right)\right)=\frac{d}{dt}\left(v_x - V_x\right)=\frac{dv_x}{dt}.
\end{equation}
It is clear that the condition (\ref{invariancia:relativaGal}) is satisfied for $v_x$ and $v_{\bar{x}}$. Trivially, it is also satisfied for the $y$ and $z$ axes. Consequently, Galilean transformations satisfy (\ref{invariancia:relativaGal}).
However, the Galilean and Newtonian models contradict the condition of the natural referent, --expression (\ref{referente:natural1})--, since there is no possibility of finding an entity with constant motion with respect to any coordinate change in the Galilean model.

\section{Ontological Test of Lorentz Transformations}

Let $X$ be a relative space, and let $X \times \mathbb{R}$ be a spacetime equipped with a Minkowski metric. The coordinate transformations that maintain this metric invariant are the Lorentz transformations:
\begin{equation}
\begin{cases}
    \bar{\mathbf{r}} = \mathbf{r} + \mathbf{v}\left[\frac{(\mathbf{r}\cdot\mathbf{v})}{v^2}(\gamma_v - 1) - t\gamma_v\right]\\
    \bar{t} = \gamma_v\left( t - \frac{\mathbf{r}\cdot \mathbf{v}}{c}\right)
\end{cases}
\end{equation}
where 
\begin{equation}
    \gamma_v = \frac{1}{\sqrt{1-\frac{v^2}{c^2}}}
\end{equation}
with $\mathbf{v} = (v_x,v_y,v_z)$ and $|\mathbf{v}| = v$.\\

Now, the velocity of an object is given by
\begin{equation}
\mathbf{u} = \frac{d\mathbf{r}}{dt}.  
\end{equation}
According to the principle of ontological invariance, it must hold that
\begin{equation}
\frac{d\mathbf{u}}{dt} = 0 \quad\Longleftrightarrow\quad \frac{d\bar{\mathbf{u}}}{dt} = 0
\label{invariancia:relativaLor}
\end{equation}
which, in terms of acceleration vector ($\displaystyle\mathbf{a} = \frac{d\mathbf{u}}{dt}$), becomes
\begin{equation}
\mathbf{a} = 0 \quad\Longleftrightarrow\quad \bar{\mathbf{a}} = 0.
\label{invariancia:relativaAcc}
\end{equation}

To demonstrate the relation (\ref{invariancia:relativaAcc}), consider that Lorentz transformations operate on the acceleration vector as follows:
\begin{equation}
    \bar{\mathbf{a}} = \frac{\mathbf{a}}{\gamma_v^2\left( 1- \frac{-\mathbf{v}\cdot\mathbf{u}}{c^2}\right)^2} - \frac{(\mathbf{a}\cdot\mathbf{v})\mathbf{v}(\gamma_v - 1)}{v^2\gamma_v^3\left( 1- \frac{-\mathbf{v}\cdot\mathbf{u}}{c^2}\right)^3} + \frac{-(\mathbf{a}\cdot\mathbf{v})\mathbf{u}}{v^2\gamma_v^2\left( 1- \frac{-\mathbf{v}\cdot\mathbf{u}}{c^2}\right)^3},
\end{equation}

\begin{equation}
    \mathbf{a} = \frac{\bar{\mathbf{a}}}{\gamma_v^2\left( 1- \frac{\mathbf{v}\cdot\bar{\mathbf{u}}}{c^2}\right)^2} - \frac{(\bar{\mathbf{a}}\cdot\mathbf{v})\mathbf{v}(\gamma_v - 1)}{v^2\gamma_v^3\left( 1- \frac{\mathbf{v}\cdot\bar{\mathbf{u}}}{c^2}\right)^3} + \frac{(\bar{\mathbf{a}}\cdot\mathbf{v})\bar{\mathbf{u}}}{v^2\gamma_v^2\left( 1- \frac{\mathbf{v}\cdot\bar{\mathbf{u}}}{c^2}\right)^3},
\end{equation}
due to this, when $\mathbf{a} = 0$, we have
\begin{equation}
     \bar{\mathbf{a}}= \frac{0}{\gamma_v^2\left( 1- \frac{-\mathbf{v}\cdot\mathbf{u}}{c^2}\right)^2} - \frac{0}{v^2\gamma_v^3\left( 1- \frac{-\mathbf{v}\cdot\mathbf{u}}{c^2}\right)^3} + \frac{0}{v^2\gamma_v^2\left( 1- \frac{-\mathbf{v}\cdot\mathbf{u}}{c^2}\right)^3}=0.
\end{equation}
And vice versa, when $\bar{\mathbf{a}} = 0$ we have
\begin{equation}
     \mathbf{a} = \frac{0}{\gamma_v^2\left( 1- \frac{\mathbf{v}\cdot\bar{\mathbf{u}}}{c^2}\right)^2} - \frac{0}{v^2\gamma_v^3\left( 1- \frac{\mathbf{v}\cdot\bar{\mathbf{u}}}{c^2}\right)^3} + \frac{0}{v^2\gamma_v^2\left( 1- \frac{\mathbf{v}\cdot\bar{\mathbf{u}}}{c^2}\right)^3}
     =0.
\end{equation}


Hence, the relation (\ref{invariancia:relativaAcc}) is demonstrated: $\bar{\mathbf{a}}=0$ if and only if $\mathbf{a} = 0$.\\

Unlike Galilean relativity, special relativity also satisfies the second principle related to the natural referent, as in this case, it is possible to find an entity whose velocity remains constant regardless of the frame of reference – in this case, $c$, the speed of light. Indeed, we have
\begin{equation}
\frac{du^i}{dt} = 0 \Longleftrightarrow \frac{du^i}{dc} = 0
\label{referente:naturalTER}
\end{equation}
which means that in this model, an object possesses a constant velocity if and only if its velocity doesn't change relative to the speed of light.\\

The relation (\ref{referente:naturalTER}) trivially holds when $c$ is constant with respect to time in any coordinate system, as the model itself restricts the reference frames to the set of inertial reference frames. Note that in Galilean relativity, this was not the case for $c$, so this condition was not fulfilled. The relation (\ref{referente:naturalTER}) is also not fulfilled when non-inertial reference frames are included in the model, such as in the case of a Rindler observer. In these cases, due to the violation of (\ref{referente:naturalTER}), ontological inconsistencies arise.\\

In conclusion, the mathematical description used in the special theory of relativity is ontologically consistent, with the peculiarity that according to the implied physical interpretation, light is the natural referent of space.\\

Next, we will present the ontological model of special relativity, conceiving motion as a change of position in spacetime $X$ relative to proper time $\tau$. In this case, the position of any object in this space is given by the 4-position:
\begin{equation}
    \textbf{x} = \left[\begin{array}{c}
         x^0(\tau) \\
         x^1(\tau) \\
         x^2(\tau) \\
          x^3(\tau)
    \end{array}\right]
\end{equation}
where component 0 corresponds to time $x^0(\tau) = ct$, and the remaining components correspond to the components of 3-position. The 4-velocity is then given by
\begin{equation}
     \textbf{U} = \frac {d\textbf{x}}{d\tau}.
\end{equation}
The principle of ontological invariance is then expressed as
\begin{equation}
     \frac{d\textbf{U}}{d\tau} = 0 \Longleftrightarrow  \frac{d\bar{\textbf{U}}}{d\tau}=0,
     \label{eq:4invar}
\end{equation}
which, in terms of 4-acceleration, would be
\begin{equation}
     \textbf{A} = 0 \Longleftrightarrow  \bar{\textbf{A}}=0.
     \label{eq:4invarAcel}
\end{equation}

This relation can be demonstrated as follows: first note that the components of any 4-vector in two inertial reference frames with relative velocity $v$ are connected by Lorentz transformations:
 \begin{equation}
     \bar{\mathbf{A}}_r = \mathbf{A}_r + \frac{\mathbf{v}}{c}\left[\frac{(\mathbf{A}_r\cdot \mathbf{v})c}{v^2}(\gamma_v - 1) - \mathbf{A}_t\gamma_v\right],
 \end{equation}

 \begin{equation}
     \mathbf{A}_r = \bar{\mathbf{A}}_r + \frac{\mathbf{v}}{c}\left[\frac{(\bar{\mathbf{A}}_r\cdot \mathbf{v})c}{v^2}(\gamma_v - 1) + \bar{\mathbf{A}}_t\gamma_v\right],
 \end{equation}

\begin{equation}
    \bar{A}_t = \gamma_v\left(A_t - \frac{\mathbf{A}_r\cdot \mathbf{v}}{c} \right),
\end{equation}

\begin{equation}
    A_t = \gamma_v\left(\bar{A}_t + \frac{\bar{\mathbf{A}}_r\cdot \mathbf{v}}{c} \right).
\end{equation}

In the above expressions, it can be seen that setting $A_t = 0$ and $\mathbf{A}_r = 0$ results in $\bar{\mathbf{A}} = 0$. Similarly, setting $\bar{A}_t = 0$ and $\bar{\mathbf{A}}_r = 0$ results in $\mathbf{A} = 0$. This proves the expression (\ref{eq:4invarAcel}).
\qed\\

On the other hand, the second ontological condition, expressed in 4-vectors, would be as follows:
\begin{equation}
\frac{du^\mu}{d\tau} = 0 \Longleftrightarrow \frac{d\left(u^\mu - c\right)}{d\tau} = 0.
\label{referente:naturalTER4}
\end{equation}
For this relation to hold, it is necessary and sufficient that $\frac{dc}{d\tau} = 0$, meaning that $c$ is also constant with respect to proper time $\tau$. This is the case because
\begin{equation}
    \frac{dc}{d\tau} = \frac{dc}{dt} \frac{dt}{d\tau} = 0 \cdot \gamma(u) = 0.
\end{equation}

However, it should be noted that $c$ is the magnitude of a 3-velocity. This raises the following question: Does a \textit{4-velocity} exist whose magnitude serves as a natural reference in the description of motion using 4-vectors?
In such a case, the following should hold:
\begin{equation}
\frac{du^\mu}{d\tau} = 0 \Longleftrightarrow \frac{d\left(u^\mu - \kappa\right)}{d\tau} = 0
\label{eq:4ref2}
\end{equation}
where $\kappa$ is the magnitude of a 4-velocity. However, the magnitude of the 4-velocity of any object is an invariant constant:
\begin{equation}
    ||\textbf{U}|| =  U^\mu U_\mu = c.
\end{equation}
Thus, $c$ turns out to be the magnitude of the 4-velocity of any object, and consequently $\kappa = c$. The relation (\ref{eq:4ref2}) becomes (\ref{referente:naturalTER4}), and we then have the following proposition.\\

\begin{proposition}
The model that describes motion using 4-vectors in Minkowski space, along with the rate of change using the total derivative $\frac{d}{d\tau}$, is ontologically consistent.
\end{proposition}

However, it is important to observe that in the 4-vector model, the velocity $c$ of the natural reference \textit{is not} the 4-velocity of light, as light does not possess a 4-velocity. In this model, $c$ is the 4-velocity of any object moving with a 3-velocity lower than that of light. In other words, any object traveling at speeds less than the speed of light ($u < c$) serves as a natural reference for the 4-vector model.
This result is interesting, as presumably, any object endowed with 4-velocity has mass. This assertion will be interesting to examine in the next section, where we will study the ontological properties of general relativity, in which mass turns out to be the cornerstone in the deformation of spacetime.

\section{Ontological Test of General Theory of Relativity}

In principle, general relativity is ontologically valid locally, as general relativistic spaces are locally Lorentzian. As demonstrated in the previous section, any Lorentzian space is ontologically consistent. What we now wish to determine is whether this ontological consistency holds in a space deformed by a relativistic metric. 

Let's consider the field equations of general relativity:
\begin{equation}
    R_{\mu \nu} - \frac{1}{2}Rg_{\mu \nu} + \Lambda g_{\mu \nu} = \frac{8\pi G}{c^4}T_{\mu \nu}
\end{equation}
where $R_{\mu\nu}$ is the Ricci curvature tensor, $R$ is the curvature scalar, $g_{\mu\nu}$ is the metric tensor, $\Lambda$ is the cosmological constant, $G$ is the Newtonian gravitational constant, $c$ is the speed of light in vacuum, and $T_{\mu \nu}$ is the energy-momentum tensor.

In the case of General Theory of Relativity (GTR), the 4-acceleration is tensorially expressed using the covariant derivative:
\begin{equation}
    A^\lambda = \frac{DU^\lambda}{d\tau} = \frac{dU^\lambda}{d\tau} + \Gamma^\lambda_{\mu\nu}U^\mu U^\nu.
\end{equation}
This expression is invariant under coordinate transformations, so when the components are zero in one system, they are also zero in another, and vice versa. Consequently,
\begin{equation}
    A^\lambda = 0 \Longleftrightarrow \bar{A}^\lambda = 0
\end{equation}
for $\lambda = 0,1,..,3$.

Since the last three components of the 4-acceleration correspond to the 3-acceleration, the first principle of ontology is also fulfilled in GTR:
\begin{equation}
    \mathbf{a} = 0 \Longleftrightarrow \bar{\mathbf{a}} = 0.
\end{equation}

In GTR, the existence of a natural referent in the description with 3-vectors is still valid, and for the description in 4-vectors, we encounter the same case of a natural referent as in special relativity. Thus, it can be seen that GTR \textit{inherits} the ontological consistency of special relativity.

In summary, we can say that up to this point, we have observed the failure of Galilean relativity in terms of ontological tests, while Einstein's special and general relativity have passed the test without issues, even when motion is modeled using 4-vectors. This analysis reveals the validity these theories exhibit when one wishes to reflect on them in terms of being and becoming, which in our opinion, is a positive outcome for philosophical contemplation.


\chapter*{Conclusions}
\addcontentsline{toc}{chapter}{Conclusions}

In this thesis, we have proposed a method for formally modeling physics and have put forth a formal ontological structure for describing the basic concepts of physics: motion, space, and time. Particularly, we have presented a novel conception that defines time as the displacement of the clock, which resolves the difficulties that the notion of simultaneity presents within the relativistic framework. This way of framing things allows for a perfectly consistent presentism that is compatible with relativistic physics, eliminating the need for eternalist approaches.

The results have also allowed us to discover the necessity of certain mathematical conditions that are linked to the ontological consistency of a physical theory. Using these conditions, we have developed consistency tests for different theories, concluding through these tests that classical physics had certain flaws, whereas special and general relativity exhibit at least the level of ontological consistency required by our model. This seems to demonstrate the usefulness of ontological studies within physics.

\chapter*{Appendix}
\addcontentsline{toc}{chapter}{Appendix}

\paragraph{Theorem \ref{teor:espaciometrico}}
The function $d$ is a metric on the space $X$.

\paragraph{Notation.} Taking into account that the set $S_{a,b}$ doesn't contain its endpoints, we will write $S_{a,b]}$ instead of $S_{a,b} \cup \{b\}$ and $S_{[a,b}$ instead of $S_{a,b} \cup \{a\}$.
\\

\begin{proof}
The distance function $d$ in a metric space $X$ must be defined on $X^2$ and take values in the real numbers, as it indeed is, and it must also satisfy the following conditions\medskip

\begin{enumerate}
    \item $d(a,b) = 0 \Leftrightarrow a = b$,
    \item $d(a,b) = d(b,a)$ (symmetry),
    \item $ d(a,c) \leq d(a,b) + d(b,c)$ (triangle inequality).
\end{enumerate}\medskip

If $a = b$, then $S_{a,b}$ is empty, and thus the displacement is $s(a,b) = 0$. Therefore, $\mu(S_{a,b}^*) = \{0\}$ and $d(a,b) = \min\{0\} = 0$, which satisfies the first condition. Symmetry directly follows from the fact that the intermediate points between $a$ and $b$ on a given path are the same regardless of the order in which the endpoints are expressed. Hence $S_{a,b} = S_{b,a}$.\\

Finally, let's prove that $d(a,b) + d(b,c) \geq d(a,c)$. Let $S_{a,c}$ be the path from $a$ to $c$ with the minimum displacement, that is, $\mu(S_{a,c}) = \min \mu(S_{a,b}^*)$. If $b \in S_{a,c}$, then $S_{a,c} = S_{a,b]} \cup S_{b,c}$ and furthermore, we have
\begin{equation}
    \mu(S_{a,b]} \cup S_{b,c}) = \mu(S_{a,b]}) + \mu(S_{b,c}) = \min\mu(S_{a,b]}^*) + \min \mu(S_{b,c}^*).
\end{equation}
Since if the paths $S_{a,b]}$ and $S_{b,c}$ were not minimal, neither would be $S_{a,c}$, contrary to our assumption. In this case, the following holds:
\begin{equation}
\min\mu(S_{a,c}^*) = \min\mu(S_{a,b]}^*) + \min\mu(S_{b,c}^*) .   
\label{eq:equ}
\end{equation}

This is $d(a,c)=d(a,b) + d(b,c)$. Now let $b'$ be a point such that $b'\notin S_{a,c}$. Then there are two possible cases: either $S_{a,b'}$ is part of another minimal path from $a$ to $c$, and thus a similar relationship to \ref{eq:equ} holds, or it is not, and in that case, the path from $a$ to $c$ passing through $b'$ would be longer:
\begin{equation}
\min\{\mu(S_{a,c}^*)\} < \min\{\mu(S_{a,b']}^*) + \mu(S_{b',c}^*)\}    
\label{eq:may}
\end{equation}
That is, $d(a,c)<d(a,b') + d(b',c)$. Combining \ref{eq:equ} and \ref{eq:may} yields the triangle inequality.
\end{proof}


\bibliographystyle{acm} 

\addcontentsline{toc}{chapter}{Bibliography}


\begin{thebibliography}{X}

\bibitem{3} Escobedo S. (2011),
  \textit{Teoría de los entes}. Con una introducción a la lógica y a la semiótica. Temacilli, México.

\bibitem{5} Pietschmann, H. (2016),
  \textit{Is General Relativity a (partial) Return of Aristotelian Physics?}, arXiv preprint arXiv: \textbf{1604.06491.}
  
    \bibitem{2} Escobedo S. (2017),
  \textit{Deducci\'on de los principios de la relatividad a partir de Arit\'oteles.}, Presentation. II Congreso de Filosof\'ia de la Ciencia. Guadalajara, México
  

  \bibitem{4} Lokajicek, M. V. (2007),
  \textit{Phenomenological and ontological models in natural science.} arXiv preprint arXiv: \textbf{0710.3225.}
  
    \bibitem{6} Dieks, D. (2006),
  \textit{The ontology of spacetime (Vol. 1).} Elsevier.
  
  \bibitem{1} Carroll, S. (2014),
  \textit{Spacetime and Geometry: An Introduction to General Relativity}, Pearson new internal edition, p. 235, .
  
  \bibitem{7} Heisenberg, Werner (1969), \textit{Der Teil und das Ganze}, München Piper, p. 41. Trad. cast. Diálogos sobre la física atómica. México, Univ. Autónoma de Puebla s/f p. 39.
  
  \bibitem {hedrich2007internal} Hedrich, Reiner (2007), \textit{The internal and external problems of string theory: A philosophical view},  Journal for General Philosophy of Science, Springer 38, 2,  261--278.
  
  \bibitem{sistema2006} Oficina Internacional de pesos y medidas (2006), \textit{El Sistema Internacional de Unidades-SI}, Centro Español de Meteorología.
  
  \bibitem{31} Escobedo S. (2016), \textit{Ontolog\'ia  del movimiento: hacia una ontolog\'ia sistem\'atica de la f\'isica moderna (partes 1 y II)}. LIX Congreso Nacional de F\'isica, México.
  
  
  \bibitem{Loka} Lokajicek, M. V. (2007), \textit{Phenomenological and ontological models in natural science}. arXiv preprint arXiv: \textbf{0710.3225.}
  
  \bibitem{Bro} Brody, T. A. (2012), \textit{The philosophy behind physics}. Springer Science \& Business Media.
  
  \bibitem{Rov} Rovelli, C. (2015), \textit{Aristotle's Physics: A Physicist's look}. Journal of the American Philosophical Association, 1(1), 23-40.
  
  
  \bibitem{Ter} Terekhovich, V. (2018), Metaphysics of the principle of least action. Studies in History and Philosophy of Science Part B: Studies in History and Philosophy of Modern Physics, 62, 189-201.

\bibitem {chaverondier2008special} Chaverondier, Bernard. (2008), \textit{Special Relativity and possible Lorentz violations consistently coexist in Aristotle space-time}, arXiv preprint arXiv: \textbf{0805.2417}, 
 
\bibitem {romero2011philosophical} Romero, Gustavo E. (2011),  \textit{Philosophical problems of space-time theories}, arXiv preprint arXiv: \textbf{1105.4376}.

\bibitem {romero2013change}  Romero, Gustavo E. (2013), \textit{From change to spacetime: an Eleatic journey}, Foundations of Science, Springer 18, 1, 139--148.

\bibitem {romero2012parmenides}  Romero, Gustavo E. (2012), \textit{Parmenides reloaded}, Foundations of Science, Springer 17, 3,  291--299.

\bibitem {solov2012physics} Solov'ev, EA. (2012), \textit{Physics and Metaphysics}, arXiv preprint arXiv: \textbf{1212.1299}.

\bibitem {caruso2015causa} Caruso, Francisco and Xavier, Roberto Moreira. (2015), \textit{Causa Efficiens versus Causa Formalis: origens da discuss\~{a}o moderna sobre a dimensionalidade do espaço},  arXiv preprint arXiv: \textbf{1505.00501}.

\bibitem {romero2017ontology} Romero, Gustavo E. (2017), \textit{On the ontology of spacetime: Substantivalism, relationism, eternalism, and emergence}, Foundations of Science, Springer 22, 1,  141--159.

\bibitem {bunge1973filosofia}  Bunge, Mario Augusto (1973), \textit{Filosofía de la física}, D. Reidel Publishing Company, Dordrecht, Holanda. 

\bibitem {bunge2011ontologia}  Bunge, Mario Augusto (2011), \textit{Ontología I: El moblaje del mundo: Volumen III. Tratado de Filosofía}, Editorial Gedisa, vol. 3.

\bibitem {Azz} Agazzi, E. (1978), \textit{Temas y problemas de la filosof{\'\i}a de la f{\'\i}sica}, Ed. Herder, Barcelona.

\bibitem{besnard2011time} Besnard, Fabien (2011), \textit{Time of philosophers, time of physicists, time of mathematicians}, arXiv preprint arXiv: \textbf{1104.4551}
\end{thebibliography}
\renewcommand*{\bibname}{Bibliography}

\end{document}